\documentclass{article}
\begin{document}\begin{center}

   {\bf Constructing Metric Gravity's N-body Non-linear Lagrangian \\ From  Iterative, Linear Algebraic Scaling Equations}
\end{center}
\begin{center}
    \bf {Kenneth Nordtvedt$^1$}
\end{center}
\begin{center}
{$^1$\small \sc 118 Sourdough Ridge, Bozeman MT 59715 USA    \small {\it E-mail address}: knordtvedt@bresnan.net}
\end{center}
\begin{center}
   {\sc Abstract}
\end{center}
\rm From a matter distribution metric gravity produces  both a set of non-linear metric field potentials $g_{\mu\nu}$ which establish the space-time arena and its local proper coordinates for all physics, and also the non-linear gravitational equations of motion for the matter in that arena.  Inspired by invariance properties of General Relativity's gravity which have been empirically confirmed to good precision --- \bf exterior effacement, interior effacement, and the time dilation and Lorentz contraction features of matter under boosts \rm --- they are here exploited to develop an iterative, linear algebraic method for obtaining the N-body non-linear gravitational Lagrangian to all orders (excluding gravitational radiation reaction effects, but see \cite{N151}\cite{N152})\footnote{Gravitational radiation reaction terms in the N-body equations of motion can not be inferred by these algebraic methods.  But once introduced, they 1) can not show spectator scaling partitions to lower order terms in the equations of motion, and 2) must be accompanied by an infinite series of higher order potential modifications which produce the required spectator scaling to enforce exterior effacement for the introduced radiation reaction term. Otherwise, gravitational radiation reaction effects in otherwise identical systems would vary if  located in different surroundings.}  To illustrate the execution of the iterative method, the complete $1/c^4$ order N-body gravitational Lagrangian is obtained for the case of Eddington's spatial metric parameter $\gamma$ equal to one\cite{Edd}.  Some iterative relationships are shown to be expressible in general form, permitting certain metric and Lagrangian potential series' coefficients to be analytically given to all orders.  In particular, the complete spatial and temporal Schwarzschild metric potentials of General Relativity are obtained. 

\section{Introduction}
An iterative, linear algebraic method is developed for obtaining the complete non-linear metric gravity Lagrangian for N bodies.  Key invariance properties, inspired by General Relativity and apparently fulfilled by present-day observations, are enforced order by order to fix the potentials and their numerical coefficients which appear in the Lagrangian and important parts of the metric field.  The method, though designed for all orders, is carried out to the $1/c^4$ order in order to illustrate exercise of the enforcements.  Some potential types have their coefficients analytically determined to all orders by the method, including the Schwarzschild metric potentials; suggesting that future work may derive analytical solutions for all coefficients to all orders. Some purposes for development of this alternative approach to metric gravity include: 1. acquiring a more efficient calculational way, perhaps even programmable on a PC?, to obtain the gravitational dynamics at high orders for more accurate application to extremely relativistic situations in astrophysics; 2. provide a confirmation or checking process for more traditional methods based on perturbative solutions of General Relativity's non-linear partial differential field equations; 3. discover whether or not an alternative to General Relativity can both fulfill the foundational invariance properties, yet can be built upon a different value for Eddington's $\gamma$ parameter\footnote{In Eddington's generalization of metric gravity, at lowest order the spatial metric potential is $-g_{ab}=\left(1+2\gamma\: U(\vec{r}\:)/c^2\right)\:\delta_{ab}$ with $U(\vec{r}\:)$ being the Newtonian potential of surrounding matter. $\gamma$ is precisely one in General Relativity.\cite{Edd}} which gives the strength of the spatial metric component's lowest order potential, important for its contribution to the leading order alteration to the global speed of light in a gravitational environment. In Appendix C the algebraic method is examined for the $\gamma \neq 1$ generalization, with preliminary results obtained.

\section{Exterior Effacement}

\bf Exterior effacement \rm means that the gravitational Lagrangian as well as metric field potentials for a general local N-body system of bodies are not altered by distant distributions of spectator bodies when the local system Lagrangian and metric field are expressed in the local proper space-time coordinates  produced by the spectator bodies' rescalings. For creating necessary (but not total) conditions for achieving exterior effacement, consideration can be restricted to the motion-independent potentials for a general N-body system of bodies surrounded asymptotically by  Minkowski space; see \cite{N151}.  The temporal metric field potential $g_{00}$, the isotropic part of the spatial metric field potential $-g_{ab}\sim\delta_{ab}$, and the N-body Lagrangian all contain non-linear,  infinite-order expansions  in powers of $1/c^2$ of motion-independent potentials with the structures:
\begin{eqnarray}
-g_{ab}(\vec{r}\:)\;&=&\;\left(1\;+\;\sum_{n,\alpha }\kappa_{n,\alpha}\;V(n, \alpha ,\vec{r}\:)\right)\;\delta_{ab}\;+\;.... \\
&=& \left(1+2\sum_i\frac{G\:m_i}{c^2\:r_i}
+\kappa_{2,1}\sum_{ij}\frac{G^2\:m_i m_j}{c^4\:r_i r_j}+\kappa_{2,2}\sum_{ij}\frac{G^2\:m_i m_j}{c^4\:r_i r_{ij}}+....\right)\;\delta_{ab}+.... \nonumber
\end{eqnarray}
\begin{eqnarray}
g_{00}(\vec{r}\:)\;&=&\;1\;+\;\sum_{n,\alpha }\xi_{n,\alpha}\;V(n, \alpha ,\vec{r}\:)\;+\;.... \\
&=&1-2\sum_i\frac{G\:m_i}{c^2\:r_i}
+\xi_{2,1}\sum_{ij}\frac{G^2\:m_i m_j}{c^4\:r_i r_j}+\xi_{2,2}\sum_{ij}\frac{G^2\:m_i m_j}{c^4\:r_i r_{ij}}+.... \nonumber
\end{eqnarray}
\begin{equation}
L\;= \sum_{n,\epsilon} \lambda _{n,\epsilon } \; U(n,\epsilon ,r_{ij} ,r_{jk} .... )+.... 
=\frac{1}{2}\sum_{ij} \frac{G\: m_i m_j}{r_{ij}} +\lambda_{1,1}\sum_{ijk}\frac{G^2\:m_i m_j m_k}{c^2\:r_{ij}r_{ik}}+.... \nonumber
\end{equation}
with $r_i=|\vec{r}-\vec{r}_i |,\;r_{ij}=|\vec{r}_j-\vec{r}_i |$, $\vec{v}_{ij}=\vec{v}_i-\vec{v}_j$, etc. The +.... are acknowledgments that the full metric field and Lagrangian expansions include motion-dependent potentials as well.  Note that the dimensionless potentials $V$ in the metric field component expansions differ from the potential energies $U$ in the N-body Lagrangian expansion; the former have a field point $\vec{r}$ and latter contain only interbody intervals $r_{ij}$. The $\alpha$ and $\epsilon$ indices run over the number of distinct potential types appearing at each order in $Gm/c^2r$.\footnote{ These $\alpha$ range numbers grow rapidly with order $n$, being 1, 2, 4, 9, 20, 48, 115, 286, respectively, $n=1\;to\;8$, for example, and similar, albeit smaller sized, escalation for the $\epsilon$ range numbers.} Beginning with known or assumed lowest order terms for these three infinite series of potentials,  the goal is to develop an iterative process which determines by linear algebraic equations the dimensionless coefficients $\kappa_{n,\alpha},\;\xi_{n,\alpha},$ and $\lambda_{n,\epsilon}$  in terms of lower order $n'<n$ coefficients so as to enforce exterior effacement and interior  properties of gravity.\footnote{The first integer $n$ in these coefficients indicates the powers of $1/c^2$ in their corresponding potentials. So the lowest order Lagrangian coefficient is $\lambda_{0,1}=1/2$ for the Newtonian potential energy.}  

Equations 1-3 are assumed to apply to a general system of bodies consisting of distant spectator bodies $s,\;s',\;s''\;....$ at large distances $R_s,\;R_{s'}\;....$ from a local system of bodies $i,\;j,\;k\;....$ in whose vicinity the spectator bodies produce rescaled space and time proper coordinates.  If the metric potentials in the expansions of Equations 1-2 are confined to the spectator body sources located at great distances $R_s$ from the local system  --- for spectator sources $s,\;s',\;s'',\;....\;r_s\rightarrow R_s,\;r_{ss'}\rightarrow R_{ss'},\;etc.,\;V(n,\;\alpha,\;\vec{r})\rightarrow \overline{V}(n,\;\alpha)$, the local metric field space and time scaling factors are produced:
\begin{eqnarray}
-g_{ab}(\vec{r}\:)&\rightarrow& -\overline{g}_{SS}\:\delta_{ab}=\left(1\;+\;\sum_{n,\alpha }\kappa_{n,\alpha}\;\overline{V}(n, \alpha )\right)\:\delta_{ab} \\ 
 g_{00}(\vec{r}\:)&\rightarrow& \overline{g}_{00}=1\;+\;\sum_{n,\alpha }\xi_{n,\alpha}\;\overline{V}(n, \alpha)
\end{eqnarray}
and which give the proper coordinates for the reexpression of local system physics; $\left(\overline{g}_{00}\right)^{1/2}\:dt=d\tau$ and $\left(-\overline{g}_{SS}\right)^{1/2}\:d\vec{r}=d\vec{\rho}$.

The potentials appearing in Equations 1-2, by various specified partitions of the body summations into spectator bodies and local bodies, become dimensionless spectator scaling factors times lower order local system N-body potentials, and this split of any $nth$ order potential generally exists in  multiple ways.  Among the nine fourth order potentials\footnote{The nine $V(4,\alpha,\:\vec{r}\:)$ potentials range from the Schwarzschild-like $V(4,1,\:\vec{r}\:)$ proportional to $m_im_jm_km_l/r_ir_jr_kr_l$ to the \it structurally linear \rm $V(4,9,\:\vec{r}\:)$ proportional to $m_im_jm_km_l/r_ir_{ij}r_{jk}r_{kl}$)  } is the potential  $V(4,4,\vec{r}\:)$ whose complete set of partitionings is 
\begin{eqnarray}
& &V(4,4,\vec{r}\:)=\sum_{i,j,k,l} \frac{G^4 \: m_i m_j m_k m_l}{c^8 \: r_{i}r_{j}r_{jk}r_{jl}}\rightarrow\left[ \sum_s \frac{G \: m_s}{c^2 \: R_s}\right]\;
\sum_{local\;i,j,k}\left(\frac{G^3\:m_i m_j m_k}{c^6\:r_i r_{ij} r_{ik}}+2\:\frac{G^3\:m_i m_j m_k}{c^6\:r_i r_{j} r_{jk}}\right) \nonumber \\
& &+\left[ \sum_s \frac{G \: m_s}{c^2 \: R_s}\right]^2
\sum_{local\;i,j}\left(\frac{G^2\:m_i m_j}{c^4\:r_i r_{j} }+2\:\frac{G^2\:m_i m_j}{c^4\:r_i r_{ij} }\right)
+\left[\left( \sum_s \frac{G \: m_s}{c^2 \: R_s}\right)^3+\sum_{ss's''}\frac{G^3\:m_s m_{s'} m_{s''}}{c^6\:R_s R_{ss'} R_{ss''}}\right]
\sum_{local\;i}\frac{G\:m_i}{c^2\:r_i} \nonumber
\end{eqnarray}
with the various spectator potential scale factors shown in $[...]$ brackets. These partitions can also be written as:\footnote{Products of spectator scaling potentials equal other spectator potentials, $\overline{V}(1,1)^3=\overline{V}(3,1), \overline{V}(2,2)\overline{V}(1,1)=\overline{V}(3,2)$, etc.}
\[
V(4,4)\rightarrow \overline{V}(1,1)\left\{V(3,3)+2V(3,2)\right\}+\overline{V}(2,1)\left\{V(2,1)+2V(2,2)\right\}+\left[\:\overline{V}(3,1)+\overline{V}(3,3)\right]V(1,1)
\]
in which the $\vec{r}$ dependence of the $V(n,\alpha,\vec{r})$ is suppressed for convenience, and will be done hereafter unless some significant ilustrative purpose demands otherwise.  

Exterior effacement requires that every local system potential of any order must receive from all the higher order potentials its specific, required total rescaling.  For the potentials appearing in the spatial and temporal metric components the total required scaling factor of an $nth$ order potential proportional to $1/r^n$ must be $\left(-\overline{g}_{SS}\right)^{1-n/2}$ and $\left(\overline{g}_{00}\right)^1 \left(-\overline{g}_{SS}\right)^{-n/2}$, respectively, so that
\begin{displaymath}
\left(-\overline{g}_{SS}\right)^{1-n/2}\;V(n,\alpha, \vec{r}\:)\;dr^2\longrightarrow V(n,\alpha, \vec{\rho}\:)\;d\rho^2\mbox{ and }\left(\overline{g}_{00}\right)^1\left(-\overline{g}_{SS}\right)^{-n/2}\;V(n,\alpha, \vec{r}\:)\;dt^2\longrightarrow V(n,\alpha, \vec{\rho}\:)\;d\tau^2
\end{displaymath}
and therefore the metric potential expansions exactly reproduce themselves when expressed in the new local proper coordinates.   Requiring this to be fulfilled at each order $n$ in spectator scaling, given it is fulfilled to order $n-1$, then yields the iterative linear algebraic equations for the expansion coefficients $\kappa_{n,\alpha}$ and $\xi_{n,\alpha}$
\begin{equation}
\sum_{\alpha} c(n,\alpha ;n',\alpha',\beta)\;\kappa_{n,\alpha}\;= \kappa_{n',\alpha'}\;\left(\frac{\partial}{\partial \overline{V}(n-n',\beta )} (-\overline{g}_{SS})^{1-n'/2}
\right)
\end{equation}
\begin{equation}
\sum_{\alpha} c(n,\alpha ;n',\alpha',\beta)\;\xi_{n,\alpha}\;= \xi_{n',\alpha'}\;\left(\frac{\partial}{\partial \overline{V}(n-n',\beta )} (\overline{g}_{00})^1(-\overline{g}_{SS})^{-n'/2}\right)
\end{equation}
with there being a different linear equation for each combination of $n'<n$, $\alpha '$, and  $\beta$.  The numbers $ c(n,\alpha ;\;n',\alpha',\beta)$ are simply the integer counts of how many ways (including possibly zero) the potential $V(n,\alpha, \vec{r}\:)$ partitions into the lower order potential $V(n',\alpha', \vec{r}\:)$ times the spectator scaling potential $\overline{V}(n-n',\beta)$.  The partial derivatives of the required total scaling factors on right hand side of Equations 5,6 simply pull out of those total scaling factors the numerical coefficients of the  indicated spectator potentials.\footnote{Because all products of spectator scaling potentials equal another single spectator scaling potential, $(\overline{g}_{00})^x(-\overline{g}_{SS})^y$ for all powers $x$ and $y$, integer or otherwise, can be expressed as a linear combination of spectator scaling potentials.}  

Once the spatial and temporal metric field components are determined up to some order, the iterative algebraic equations for the motion-independent Lagrangian potentials can be constructed to a corresponding order.  The Lagrangian potentials' total scaling factors required for exterior effacement are $(\overline{g}_{00})^{1/2}(-\overline{g}_{SS})^{-(n+1)/2}$ ($U(n,\epsilon)$ is proportional to $1/r^{n+1}$) so that
\[
(\overline{g}_{00})^{1/2}(-\overline{g}_{SS})^{-(n+1)/2}\;U(n,\epsilon ,r_{ij} ,r_{jk} .... )\;dt \longrightarrow U(n,\epsilon ,\rho_{ij} ,\rho_{jk} .... )\;d\tau
\]
The resulting set of iterative linear algebraic equations for the Lagrangian coefficients $\lambda_{n,\epsilon}$ are then
\begin{equation}
\sum_{\epsilon} c^*(n,\epsilon ;n',\epsilon',\beta)\;\lambda_{n,\epsilon}\;=\lambda_{n',\epsilon'} \;\left(\frac{\partial}{\partial \overline{V}(n-n',\beta )} (\overline{g}_{00})^{1/2}(-\overline{g}_{SS})^{-n'/2}
\right)
\end{equation}
for each $n'<n$, each type $\epsilon'$, and each type $\beta$. The numbers  $c^*(n,\epsilon ;n',\epsilon',\beta)$ are now the integer counts of how many ways (including possibly zero) the Lagrangian potential $U(n,\epsilon ,r_{ij} ,r_{jk} .... )$ partitions into the lower order potential $U(n',\epsilon' ,r_{ij} ,r_{jk} .... )$ times spectator scaling potential $\overline{V}(n-n',\beta)$.  After fixing the $n$th order coefficients $\kappa_{n,\alpha},\;\xi_{n,\alpha}$ and $\lambda_{n,\epsilon}$, they are never revisited again for adjustment --- their true values have been found and fixed. The chore of better enforcing exterior effacement then moves on to finding the next higher order coefficients.  An occasionally useful tool relating the metric potential component expansions given by Equations 1,2 to the Lagrangian expansion given by Equation 3 is to take the  test particle limit of the latter and requiring that limit to match the geodesic Lagrangian for test particles given by:
\begin{equation}
\frac{L}{m}\;dt = -c^2\:\left(g_{\mu\nu}\:\frac{dx^{\mu}}{dt}\frac{dx^{\nu}}{dt}\right)^{1/2}\;dt
\end{equation}

\subsection{Example: Lowest Order Exterior Effacement}

As example of exterior effacement, its lowest order manifestation can be exhibited by dividing the Newtonian potential summation into its local body contributions and its contribution from all the distant spectator bodies $s$, $\overline{V}(1,1)\equiv U_S=\sum Gm_s/c^2R_s$, and so to linear order $(1-U_S)dt\approx d\tau,\;(1+U_S)dr\approx d\rho$:
\[
\sum_i\frac{G\:m_i}{c^2\:r_i}\rightarrow \sum_{local\;i}\frac{G\:m_i}{c^2\:r_i}+U_S,\;\;\sum_{ij}\frac{G\:m_im_j}{c^2\:r_{ij}}\rightarrow 2U_S\sum_{local\;i}m_i,
\;\;\;\sum_{ijk}\frac{G^2\:m_i m_j m_k}{c^2\:r_{ij}r_{ik}}\rightarrow 2U_S\sum_{local\;ij}\frac{G\:m_im_j}{r_{ij}} 
\]
and then the lowest order scalings of the Lagrangian expansion given by Equation 3 (plus the bodies' rest energy Lagrangian) produce
\begin{eqnarray}
& &\left(-\sum_{i}m_ic^2 + \frac{1}{2}\sum_{ij}\frac{G\:m_im_j}{r_{ij}} +\lambda_{1,1}\sum_{ijk}\frac{G^2\:m_i m_j m_k}{c^2\:r_{ij}r_{ik}}\right)\; dt \longrightarrow \nonumber \\
& &\hspace{1.0in} \left(-\left(1-U_S\right)\:\sum_{local\;i}m_ic^2+ \left(\frac{1}{2}\left(1+U_S\right)+2\:\lambda_{1,1}\:U_S\right)\sum_{local\:ij} \frac{G\: m_i m_j}{\rho_{ij}}\right)\;dt \nonumber
\end{eqnarray}
For $\lambda_{1,1}=-1/2$ an overall $1-U_S$ factors out and combines with $dt$ to produce the local proper $d\tau$.  The spectator bodies are effaced to order $U_S$ from the local physics.  This means, in particular, that Newton's $G$ is at this lowest order locally unaffected by distribution of distant matter.\cite{N84}

 
\section{The Lagrangian's Motion-Dependent Potentials}
The overall goal is to develop the complete Lagrangian expansion for N-bodies including the motion-dependent terms:
\begin{eqnarray}
L&=&-\sum_i m_i c^2 \left(1-\frac{1}{2}\frac{v_i^2}{c^2}-\frac{1}{8}\frac{v_i^4}{c^4}+....\right) \\
& &+\sum_{n,\epsilon} U\left(n,\epsilon ,r_{ij} ,r_{jk} .... \right)\;
\left\{\lambda_{n,\epsilon}+\sum_{\tilde{n},\tilde{\epsilon}}\lambda_{n,\epsilon,\tilde{n},\tilde{\epsilon}}\;
F\left(n,\epsilon;\;\tilde{n},\tilde{\epsilon},\left[\frac{dr...}{c\:dt}\right]^{2\tilde{n}},\hat{r}_{ij}....\right)\right\} \nonumber
\end{eqnarray}
The first line of Equation 10 is the kinetic Lagrangian expansion for the mass elements, $-m_i c^2 \sqrt{1-v_i^2/c^2}$. In the second line are the motion-dependent dressings for each of the motion-independent potentials.  The  functions $F\left(n,\epsilon;\;\tilde{n},\tilde{\epsilon};\;\left[\frac{dr...}{c\:dt}\right]^{2\tilde{n}},\hat{r}_{ij}....\right)$ are dimensionless, and their $\left[\frac{dr...}{c\:dt}\right]^{2\tilde{n}}$ factors indicate $2\tilde{n}$ time derivatives of mass positions, creating velocity dependence, acceleration dependence, etc. The $\tilde{\epsilon}$ label ranges over all the different forms that these motion-dependent potentials can take from  $2\tilde{n}$ time derivatives acting on positions and ways scalar products are formed from the vectors $\vec{r}_{ij},\;\vec{v}_{jk}$, etc., given the $n,\epsilon,\tilde{n}$ specifications.\footnote{For example: with $\tilde{n}=2$,$\left[\frac{dr...}{c\:dt}\right]^4$ could be proportional to $(v/c)^4,\;(v/c)^2ar/c^2,\;a^2r^2/c^4,\;\dot{a}r^2v/c^4$, or $\ddot{a}r^3/c^4$.}  The $\lambda_{n,\epsilon,\;\tilde{n},\tilde{\epsilon}}$ becme additional numerical coefficients requiring their determination by enforcing the invariance principles stated previously.

\subsection{Time Dilation in Boosted Systems}
\bf Time dilation \rm invariance means that any motions of a gravitational dynamical system must also be possible with respect to the dilated time variable of special relativity if the dynamical system is viewed from a moving frame of reference.  If a planar but otherwise general N-body system is given a boost $\vec{w}$ perpendicular to its planar dynamics, its kinetic Lagrangians $(1-v_i^2/c^2)^{1/2}$ are invariant under transformation to the dilated time variable $d\tau^*=\sqrt{1-w^2/c^2}\:dt$  with the planar velocities $\vec{u}_i=d\vec{r}_i/d\tau^*=\vec{v}_i\:(1-w^2/c^2)^{-1/2}$ being defined in terms of the dilated time.  Similarly,  each  Lagrangian potential rescaled by a boost perpendicular to planar dynamics must then have the analagous scaling property; with all $\vec{w}\cdot\vec{v}_i=0= \vec{w}\cdot\vec{r}_{ij}$
\begin{displaymath}
L(\vec{r}_{ij},\vec{r}_{ik}...., \vec{v}_i+\vec{w}....)\;dt= L(\vec{r}_{ij},\vec{r}_{ik}...., \vec{u}_i....)\;\left(\sqrt{1-w^2/c^2}\;dt\right)
\end{displaymath}
This requirement produces additional iterative linear algebraic equations for the coefficients of the motion-dependent Lagrangian potentials:
\begin{equation}
\sum_{\tilde{\epsilon}}c^{**}(n,\epsilon,\tilde{n},\tilde{\epsilon},\tilde{n}',\tilde{\epsilon}')\; \lambda_{n,\epsilon,\tilde{n},\tilde{\epsilon}} =
\lambda_{n,\epsilon,\tilde{n}',\tilde{\epsilon}'}\;\frac{\partial}{\partial\; (w/c)^{2(\tilde{n}-\tilde{n}')}} \left(1-w^2/c^2\right)^{(1/2-\tilde{n}')}
\end{equation}
with $c^{**}(n,\epsilon,\tilde{n},\tilde{\epsilon},\tilde{n}',\tilde{\epsilon}')$ being the coefficient (including possibly zero) of $(w/c)^{2(\tilde{n}-\tilde{n}')}$ in the expansion of $\left(1-w^2/c^2\right)^{(1/2-\tilde{n}')} $ when the perpendicular boost is made to all body velocities in the parent potential $U(n,\epsilon)F\left(n,\epsilon;\;\tilde{n},\tilde{\epsilon};\;\left[\frac{dr...}{c\:dt}\right]^{2\tilde{n}},\hat{r}_{ij}....\right)$, for each $\tilde{n},\;\tilde{\epsilon},\;\tilde{n}'<\tilde{n},\;\tilde{\epsilon}'$

\subsection{Lorentz Contraction of Boosted Systems}
One of the requirements of \bf Lorentz contraction \rm invariance is that any static configuration of matter requires the Lorentz-contracted version of that system to be the corresponding, internally stationary state when viewed from another velocity frame.  Consider a configuration of bodies at rest held in a static lattice by a number of interaction types $\epsilon$ (including gravity) with potentials $U(\epsilon,r_{ij}....)$  Then 
\begin{displaymath}
\sum_{\epsilon} \vec{\nabla}_a U(\epsilon,r_{ij}....)=0\;\;\;\;\mbox{for all  body sites}\;\vec{r}_a
\end{displaymath}
The same lattice in constant motion $\vec{w}$ has its interbody distances Lorentz contracted such that
\begin{equation}
r_{ij}'=r_{ij}\:\sqrt{1-(\vec{w}\cdot\hat{r}_{ij})^2/c^2}\;\;\;\mbox{and}\;\;\;\left(\vec{w}\cdot\hat{r}^{\:'}_{ij}\right)^2=\left(\vec{w}\cdot\hat{r}_{ij}\right)^2
\left(\frac{1-w^2/c^2}{1-(\vec{w}\cdot\hat{r}_{ij})^2/c^2}\right)
\end{equation}
and each static potential type in the Lagrangian is altered by acquiring an infinite series of velocity $\vec{w}$-dependent supplements:
\begin{displaymath}
 U(\epsilon,r_{ij}....)\longrightarrow U(\epsilon,r_{ij}'....)\;\left(1+F(\epsilon,\;w^2/c^2, (\vec{w}\cdot\hat{r}^{\:'}_{ij})^2/c^2....)\right)
\end{displaymath}
But the individual bodies in the moving lattice must still all be free of acceleration. For this to be true for the general boost, the modified motion-dependent potentials  must be free of $\vec{w}\cdot\hat{r}_{ij}$ terms to all orders and for each potential type. For potential proportional to product of reciprocal interbody distances, $1/[r_{ij}.....r_{yz}]$, each Lorentz contracted according to Equation 12, the explicit motion-dependent factors must take the form  $1+F(\epsilon,\;w^2/c^2, (\vec{w}\cdot\hat{r}^{\:'}_{ij})^2/c^2....)=\sqrt{1-(\vec{w}\cdot\hat{r}_{ij})^2/c^2}\;....\;\sqrt{1-(\vec{w}\cdot\hat{r}_{yz})^2/c^2}\; \sqrt{1-w^2/c^2}$.  Imposing these conditions on each of the motion-independent gravitational potentials $U(\epsilon,r_{ij},....)$ then establishes linear algebraic conditions among their associated motion-dependent supplements $\lambda_{n,\epsilon,\tilde{n},\tilde{\epsilon}}\;
F\left(n,\epsilon,\tilde{n},\tilde{\epsilon},
\left[\frac{d\:r..}{c\:dt}\right]^{2\tilde{n}},\hat{r}_{ij}....\right)$ in the gravitational N-body Lagrangian.

\section{Interior Effacement}
The \bf Interior Effacement \rm property of gravity requires that composite bodies in the spherical limit and neglecting tidal interactions are each characterized in the N-body Lagrangian by only a single "mass" attribute ---  total mass-energy --- (and their angular momenta if rotating relative to local inertial space).  A single composite gaseous body's mass-energy to $1/c^4$ order is
\begin{eqnarray}
& &M = \frac{1}{c^2}\left(\sum_{i} \vec{v}_i\cdot \frac{\partial L}{\partial\vec{v}_i}-L+....\right)=\sum_{i\;in\;body} m_i\;\left(1 +\frac{1}{2\:c^2} \left[u_i^2 -\sum_{j\;in\;body} \frac{G\:m_j}{\rho_{ij}}\right] \right.  \\
& & + \left. \frac{1}{c^4}\left[\frac{3}{8}\:u_i^4 + \sum_{jk\;in\;body}\frac{G^2\:m_j m_k}{2\:\rho_{ij}\rho_{ik}}+\sum_{j\;in\;body} \frac{G\:m_j}{4\:\rho_{ij}}
\left(3\:u_{ij}^2 - \vec{u}_i \cdot \vec{u}_j - \vec{u}_i \cdot \hat{r}_{ij}\hat{r}_{ij}\cdot \vec{u}_j \right)\right]\right) +... \nonumber
\end{eqnarray}
in which the variables $\vec{\rho}_i$ and $\vec{u}_j$... for the composite body's mass elements are  the proper coordinates at the body's location, rescaled not only by any distant spectator masses but also by any other local system bodies. Then the total energy of a collection of composite bodies takes the identical form as Equation 13 but expressed in terms of the composite bodies' mass parameters $M_I$ as given by the internal expansions from the same Equation 13, their interbody separations $R_{IJ}$, and body velocites $\vec{W}_I$, etc.
\begin{eqnarray}
& &E = \left(\sum_i \vec{v}_i\cdot \frac{\partial L}{\partial\vec{v}_i}-L+....\right)=\sum_I M_I\;c^2\;\left(1 +\frac{1}{2\:c^2} \left[W_I^2 -\sum_J \frac{G\:M_J}{R_{IJ}}\right] \right. \nonumber \\
& & + \left. \frac{1}{c^4}\left[\frac{3}{8}\:W_I^4 + \frac{1}{2}\sum_{JK}\frac{G^2\:M_j M_k}{R_{IJ}R_{IK}}+\frac{1}{4}\sum_J \frac{G\:M_j}{R_{IJ}}
\left(3\:W_{IJ}^2 - \vec{W}_I \cdot \vec{W}_J - \vec{W}_I \cdot \hat{R}_{IJ}\hat{R}_{IJ}\cdot \vec{W}_J \right)\right]\right) +... \nonumber
\end{eqnarray}

The potentials which appear in the basic expansions for metric field components $g_{00}$ and $g_{SS}$ and the Lagrangian, Equations 1-3, are expressed in terms of "mass elements" $m_i$ as sources.  Interior effacement also requires that these same expansions can substitute composite body $M_I$ source strengths for the mass element factors $m_i$ in the limit that the composite bodies are considered in their spherical limits.  To enforce this will involve working with "internalizations" of some of the interbody links in the various potentials.  Here is an example of such internalizations ($int$).
\begin{displaymath}
\sum_{ijklq}\frac{G^5\:m_im_jm_km_lm_q}{c^{10}\:r_ir_{ij}r_{jk}r_{kl}r_{kq}}\rightarrow
\sum_{ii'kk'k''}\frac{G}{c^2\:R_I}\frac{G\overbrace{m_im_{i'}}^{int}}{c^2\:r_{ii'}}
\frac{G}{c^2\:R_{IK}}\frac{G^2\:\overbrace{m_km_{k'}m_{k''}}^{int}}{c^4\:r_{kk'}r_{kk''}}
\end{displaymath}
A 5th order metric potential becomes part of a second order potential of two composite bodies, $I$ and $K$, with one body having its internal Newtonian gravitational energy becoming part of its source strength (sums over $i$ and $i'$ within body $I$), and for the other body $K$ the internalized next order gravitational potential energy (sums over $k$, $k'$, and $k''$ within body $K$) contributes to its source strength $M_K$.  

More specifically, for a pair of composite bodies at rest, the Lagrangian-based energy must add up to
\begin{displaymath}
E = M_A c^2 + M_B c^2 - \frac{G\:M_A M_B}{R_{AB}}+\frac{G^2\;M_AM_B(M_A+M_B)}{2\:c^2\:R_{AB}^2}+....
\end{displaymath}
with $M_A$ and $M_B$ being the total mass-energies of each composite body as given by Equation 13 when expressed in each body's proper coordinates.  Body $A$ consists of a self-gravitating gas of mass elements with labels $a,\;a',\;...$, and correspondingly for body B.  Working at the $1/R_{AB}$ level Newtonian potential energy, the product of the composed body masses can then be expressed as a series of contributions, each of which must be generated from specific energy contributions from the basic N-body Lagrangian for mass elements. The Newtonian potential energy is also characterized by its single $1/R_{AB}$ factor of inter-body (non-internaliized) distance.  The product of composite body masses is a sum of individual contributions which to the $1/c^4$ order are: 
\begin{eqnarray}
& &M_A M_B = \sum_a m_a \sum_b m_b 
+\frac{1}{2c^2}\left(\sum_a m_a \left(\sum_b m_b u_b^2- \sum_{bb'}\frac{G\:m_bm_{b'}}{\rho_{bb'}}\right)+ \sum_b m_b \left(\sum_a m_a u_a^2
-\sum_{aa'}\frac{G\:m_am_{a'}}{\rho_{aa'}}\right)\right)\nonumber \\
& &+\frac{1}{4c^4}\left(\sum_a m_a u_a^2\sum_b m_b u_b^2 + \sum_{aa'}\frac{G\:m_am_{a'}}{\rho_{aa'}}\sum_{bb'}\frac{G\:m_bm_{b'}}{\rho_{bb'}} 
-\sum_a m_a u_a^2 \sum_{bb'}\frac{G\:m_bm_{b'}}{\rho_{bb'}} - \sum_b m_b u_b^2 \sum_{aa'}\frac{G\:m_am_{a'}}{\rho_{aa'}}\right) \nonumber \\
& &+\frac{1}{c^4}\sum_a m_a\left( \frac{3}{8}\sum_b m_b u_b^4 + \frac{1}{2}\sum_{bb'b''}\frac{G^2\:m_bm_{b'}m_{b''}}{\rho_{bb'}\rho_{bb''}}
+\frac{1}{4}\sum_{bb'} \frac{G\:m_b m_{b'}}{\rho_{bb'}}
\left(3\:u_{bb'}^2 - \vec{u}_b \cdot \vec{u}_{b'} - \vec{u}_b \cdot \hat{\rho}_{bb'}\hat{\rho}_{bb'}\cdot \vec{u}_{b'} \right)\right)  \nonumber \\
& &+\frac{1}{c^4}\sum_b m_b\left( \frac{3}{8}\sum_a m_a u_a^4 + \frac{1}{2}\sum_{aa'a''}\frac{G^2\:m_am_{a'}m_{a''}}{\rho_{aa'}\rho_{aa''}}
+\frac{1}{4}\sum_{aa'} \frac{G\:m_a m_{a'}}{\rho_{aa'}}
\left(3\:u_{aa'}^2 - \vec{u}_a \cdot \vec{u}_{a'} - \vec{u}_a \cdot \hat{\rho}_{aa'}\hat{\rho}_{aa'}\cdot \vec{u}_{a'} \right)\right)  \nonumber
\end{eqnarray}
so will receive contributions from a number of different types of Lagrangian potentials.  At the $1/c^4$ order, there is also a contribution to the $1/R_{AB}^2$ order potential energy between the bodies which needs to be correctly generated from the fundamental N-body Lagrangian:
\begin{eqnarray}
& &E= \frac{G^2}{2c^2\;R_{AB}^2}\left\{\sum_am_a\;\sum_bm_b\;\left(\sum_{a'}m_{a'}+\sum_{b'}m_{b'}   \right) \right. \nonumber \\
& &\left.+\frac{1}{2c^2}\left(\sum_am_au_a^2-\sum_{aa'}\frac{G\;m_am_{a'}}{\rho_{aa'}}\right)\sum_bm_b\left(\sum_{b'}m_{b'}+2\sum_{a''}m_{a''}\right) \right\}
+\;\mbox{same for}\; A\leftrightarrow B \nonumber
\end{eqnarray}

Since each composite body is not in isolation, but is in the vicinity of the other, each body ($A$ or $B$) acts as a spectator to rescale the proper space and time coordinates at the location of the other body ($B$ or $A$).  This produces a variety of rescaling factors proportional to $1/R_{AB}$ (and higher inverse powers of $R_{AB}$) which act on the interior energy contributions to each composite body's mass-energy.  The rescalings of the basic variables in the energy terms needed for present purposes are:
\begin{eqnarray}
& &v_a^2 \longrightarrow \overline{g}_{00}(-\overline{g}_{SS})^{-1} u_a^2=u_a^2 \;\left[1-2\frac{G\:(M_B+M_B^*)}{c^2\:R_{AB}}+....\right] +\hspace{.1in} \mbox{same for}\;v_b^2\;\mbox{in presence of body}\;A \nonumber \\
& &\frac{1}{r_{aa'}} \longrightarrow (-\overline{g}_{SS})^{1/2}\frac{1}{\rho_{aa'}}=\frac{1}{\rho_{aa'}}\;\left[1+\frac{G\:M_B}{c^2\:R_{AB}}+....\right] +\hspace{.1in}
 \mbox{same for}\;\frac{1}{r_{bb'}}\;\mbox{in presence of body}\;A \nonumber 
\end{eqnarray}
with $\rho_{aa'}$ being proper distance between mass elements $a$ and $a'$ in body $A$, $\vec{u}_b$ being proper velocity of mass element $b$ in body $B$, etc.
Note the appearance of 
\begin{displaymath}
M_B^*=M_B+\frac{1}{c^2}\left(\sum_b m_bu_b^2-\sum_{bb'}\frac{G\:m_bm_{b'}}{2\:\rho_{bb'}}+\mbox{higher order virial terms}\;\right) +....
\end{displaymath}
which is the composite body mass parameter which appears in the $V(1,1,\vec{r}\:)$ or $\overline{V}(1,1)$ potentials in $g_{00}(\vec{r}\:)$ or $\overline{g}_{00}$.  It includes the composite body's virial in addition to its mass-energy content.\cite{N68}  The total rest energy expressions given by Equation 13 for the two composite bodies can now be reexpressed in terms of their proper coordinates:  
\begin{eqnarray}
& &E= \sum_a m_a c^2 +\frac{1}{2}\sum_a m_a u_a^2\:\left[1-2\frac{G\:(M_B+M_B^*)}{c^2\:R_{AB}}\right. \nonumber \\
& &\;\;\;\;\left.+\frac{G^2\left(\xi_{2,1}M_B^{*2}+4M_BM_B^*+\left[4-\kappa_{2,1}\right]M_B^2+\xi_{2,1}M_B^*M_A^*-\kappa_{2,2}M_BM_A\right)}{c^4\:R_{AB}^2}\right]\nonumber \\
& & -\frac{1}{2}\sum_{aa'}\frac{G\:m_a m_{a'}}{\rho_{aa'}}\;
\left[1+\frac{G\:M_B}{c^2R_{AB}}+\frac{G^2\:\left(\left[\kappa_{2,1}-1\right]M_B^2+\kappa_{2,2}M_BM_A\right)}{2c^4\:R_{AB}^2}\right] \nonumber \\
& &+\frac{3}{8c^2} \sum_a m_a u_a^4\:\left[1-4\frac{G\:(M_B+M_B^*)}{c^2R_{AB}}\right]+\frac{1}{2c^2} \sum_{aa'a''}\frac{G^2\:m_a m_{a'}m_{a''}}{\rho_{aa'}\rho_{aa''}}\:\left[1+2\frac{G\:M_B}{c^2R_{AB}}\right] \nonumber \\
& &+\frac{1}{4c^2}\sum_{aa'} \frac{G\:m_a m_{a'}}{\rho_{aa'}}\left(3\:u_{aa'}^2 - \vec{u}_a \cdot \vec{u}_{a'} - \vec{u}_a \cdot \hat{\rho}_{aa'}\hat{\rho}_{aa'}\cdot \vec{u}_{a'} \right)\left[1-\frac{G\:(M_B+2M_B^*)}{c^2R_{AB}}\right] \nonumber \\
& &+\frac{1}{c^2}\sum_b\frac{G\:m_b}{R_{AB}}\sum_{aa'}\frac{G\:m_am_{a'}}{\rho_{aa'}}\left[1+\frac{G\:M_B}{c^2\:R_{AB}}\right]\nonumber \\
& &+\frac{3}{2c^2}\sum_{b}\frac{G\:m_b}{R_{AB}}\sum_a m_a u_a^2\left[1-2\frac{G\:(M_B+M_B^*)}{c^2R_{AB}}\right]  +\;\; A\leftrightarrow B \nonumber
\end{eqnarray}
with $M_B$ and $M_B^*$ expanded to whatever order in $1/c^2$ needed to reach $1/c^4$ total order.  These rescalings of the system's different mass-energy contributions into their forms using proper coordinates generate a number of system energy contributions proportional to $1/R_{AB}$ and $1/R_{AB}^2$.  And the basic N-body system Lagrangian-based expansion for system energy up to $1/c^4$ also explicitly yields energy contributions proportional to these same powers of $1/R_{AB}$.  The sum of energies proportional to $1/R_{AB}$ in particular, from rescaling and from explicit contribution, are then required to yield  the Newtonian potential energy  with strength  $-G\:M_A M_B$ with $M_A$ and $M_B$ expanded to the order  needed  to include all $1/c^4$ order contributions. An observationally important consequence of interior effacement is that gravitational mass remains equal to inertial mass for composite celestial bodies with significant gravitational binding energies.\cite{N68,N68b} This has been confirmed to good precision from 45 years of lunar laser ranging.\cite{LLR}  While LLR provides good measurement of the $1/c^2$ contributions to the $M_AM_B$ values for the Sun's Newtonian couplings with Earth and Moon, the $1/c^4$ order contributions to those interactions are still beyond LLR's experimental measure by about a factor of 50.  Observations of binary pulsar systems, however, present opportunities to test their Newtonian coupling strengths at the $1/c^4$ level.

\section {The Metric and Lagrangian Expansions}
Since iterative algebraic equations for potential coefficients are the primary tool for enforcing exterior effacement, the process needs the lowest order potentials as starting points. As shown in Equations 1-3, the Lagrangian expansion is assumed to begin with (negative of) the  Newtonian N-body potential energy, while the temporal metric potential expansion correspondingly begins with minus twice the Newtonian gravitational N-body potential divided by $c^2$, so that the test body geodesic equation of motion reproduces Newtonian physics in first approximation.  We assume the (negative of) spatial metric potential expansion begins with twice the Newtonian gravitational N-body potential divided by $c^2$.\footnote{Evidence for value of $\kappa_{1,1}=2$ is obtained from time delay observations of radio signals from solar system spacecrafts whose line of sights during their outward trajectories passed close to the Sun; they find $\kappa_{1,1}=2$ with precision of couple parts in $10^5$\cite{Ber}; and General Relativity theory produces exactly a $\kappa_{1,1}=2$ value.} 

At the next (first non-linear) order there are two metric field potentials and a single Lagrangian potential, each with one order of spectator scaling partitions,:
\begin{eqnarray}
& & V(2,1)=\sum_{ij}\frac{G^2\:m_im_j}{c^4\;r_ir_j}\longrightarrow 2\overline{V}(1,1)\;\sum_i \frac{G\:m_i}{c^2\:r_i}= 2 \overline{V}(1,1)\;V(1,1) \nonumber \\
& & V(2,2)= \sum_{ij}\frac{G^2\:m_im_j}{c^4\;r_ir_{ij}}\longrightarrow\overline{V}(1,1)\;V(1,1) \hspace{.4in}
 U(1,1)=\sum_{ijk}\frac{G^2\:m_im_jm_k}{c^2\:r_{ij}r_{ik}}\longrightarrow 2\overline{V}(1,1)\;U(0,1) \nonumber
\end{eqnarray}
These metric potential scaling partitions yield the three algebraic equations for metric and Lagrangian potential coefficients:
\begin{eqnarray}
& & 2\xi_{2,1}+\xi_{2,2}= \xi_{1,1}\frac{\partial}{\partial\:\overline{V}(1,1)}(\overline{g}_{00})(-\overline{g}_{SS})^{-1/2}=6 \nonumber \\
& & 2\kappa_{2,1}+\kappa_{2,2}=\kappa_{1,1}\frac{\partial}{\partial\:\overline{V}(1,1)}(-\overline{g}_{SS})^{1/2}=2 \nonumber \\
& &2\lambda_{1,1}=\lambda_{0,1}\frac{\partial}{\partial\:\overline{V}(1,1)}(\overline{g}_{00})^{1/2}(-\overline{g}_{SS})^{-1/2}=-1 \nonumber
\end{eqnarray}
The third equation fixes the coefficient $\lambda_{1,1}=-1/2$ as was previously obtained in Section 2.1.  Taking the single test particle limit of the Lagrangian and equating it to a geodesic equation, Equation 9, then fixes the temporal metric potential coefficients $\xi_{2,1}=\xi_{2,2}=2$. The separate fixes for $\kappa_{2,\alpha}$values remain to be made.

\subsection{The Spatial Metric Potential Expansion}
The fixing of the spatial metric coefficients $\kappa_{2,\alpha}$ requires a combined use of exterior effacement and interior effacement. There are two $1/c^4$ order motion-independent potentials in the N-body Lagrangian, with each having multiple scaling partitions:
\begin{eqnarray}
U(2,1)=\sum_{ijkl}\frac{G^3\:m_im_jm_km_l}{c^4\:r_{ij}r_{ik}r_{il}}&\longrightarrow& 3\overline{V}(1,1)\;U(1,1)+3\overline{V}(2,1)\;U(0,1) \nonumber \\
U(2,2)=\sum_{ijkl}\frac{G^3\:m_im_jm_km_l}{c^4\:r_{ij}r_{jk}r_{kl}}&\longrightarrow&  2\overline{V}(1,1)\;U(1,1)+\left[\overline{V}(2,1)+2\overline{V}(2,2)\right]\;U(0,1)
\nonumber \end{eqnarray}
These partitions support three algebraic equations from Equation 8 for $\lambda(2,1)$ and $\lambda(2,2)$:
\begin{eqnarray}
3\lambda_{2,1}+2\lambda_{2,2}&=&\lambda_{1,1}\;\frac{\partial}{\partial \overline{V}(1,1)}\left((\overline{g}_{00})^{1/2}(-\overline{g}_{SS})^{-1}\right)=3/2 \nonumber \\
3\lambda_{2,1}+\lambda_{2,2}&=&\lambda_{0,1}\;\frac{\partial}{\partial \overline{V}(2,1)}\left((\overline{g}_{00})^{1/2}(-\overline{g}_{SS})^{-1/2}\right)=\left(6-\kappa_{2,1}\right)/4 \nonumber \\
2\lambda_{2,2}&=&\lambda_{0,1}\frac{\partial}{\partial \overline{V}(2,2)}\left((\overline{g}_{00})^{1/2}(-\overline{g}_{SS})^{-1/2}\right)=\left(2-\kappa_{2,2}\right)/4
\nonumber \end{eqnarray}
For the case of two composite bodies A and B at rest and separated by $R$, the Lagrangian potential $\lambda_{2,2}\:U(2,2)$ includes a partially internalized $1/R$  energy contribution:
\[
-\lambda_{2,2}\:U(2,2)\longrightarrow-2\lambda_{2,2}\;\frac{G}{R_{AB}}\sum_{aa'}\frac{G\;m_am_{a'}}{c^2\:r_{aa'}}\sum_{bb'}\frac{G\;m_bm_{b'}}{c^2\:r_{bb'}}
\]
while interior effacement requires the total coefficient of such form to be $-1/4$. But there are also energy contributions of this same form from the reexpressions of the bodies' Newtonian gravitational energies into the proper spatial coordinates at each body:
\begin{displaymath}
-\frac{1}{2}\left(\sum_{aa'}\frac{G\;m_am_{a'}}{c^2\:r_{aa'}}+\sum_{bb'}\frac{G\;m_bm_{b'}}{c^2\:r_{bb'}}\right)\rightarrow -\frac{1}{2}\sum_{aa'}\frac{G\;m_am_{a'}}{c^2\:\rho_{aa'}}\left[1+\frac{G\:M_B}{c^2\:R_{AB}}\right]-\frac{1}{2}\sum_{bb'}\frac{G\;m_bm_{b'}}{c^2\:\rho_{bb'}}\left[1+\frac{G\:M_A}{c^2\:R_{AB}}\right]
\end{displaymath}
The coefficient of the Newtonian gravitational energy to $M_B$ in the spatial metric component is $\kappa_{2,2}\:/2$.  So altogether interior effacement requires $\lambda_{2,2}=(1-2\kappa_{2,2})/8$.  With the previous equation for $\lambda_{2,2}$ obtained from an exterior effacement condition, there is:
\begin{equation}
\lambda_{2,2}=\left(2-\kappa_{2,2}\right)/8=(1-2\kappa_{2,2})/8
\end{equation}     
with solution $\kappa_{2,2}=-1$ and $\kappa_{2,1}=3/2,\;\lambda_{2,2}=3/8,\;\lambda_{2,1}=1/4$.  The composite body mass parameter $M$ appearing at lowest order in the spatial metric component expansion includes just its Newtonian gravitational energy contribution $-G\sum_{aa'}m_am_{a'}/2c^2\:r_{aa'}$ without anything further, unlike the situation in the temporal metric component.  This must then be true at all orders of appearance of the composite bodies' $M_i$ in the spatial metric since the lowest order potential in this expansion receives spectator rescaling from the higher order potentials whenever a distant distribution of mass is present.

The next order potentials in the spatial metric expansion are examined to check whether they indeed fulfill interior effacement as well?  Exterior effacement enforcement leads to the algebraic equations
\begin{eqnarray}
& &3\kappa_{3,1}+\kappa_{3,2} =0=\kappa_{3,2}+2\kappa_{3,3}+\kappa_{3,4} \nonumber \\
& &3\kappa_{3,1}+\kappa_{3,2}+\kappa_{3,3}+\kappa_{3,4}=\kappa_{1,1}\;\frac{\partial}{\partial \overline{V}(2,1)}(-\overline{g}_{SS})^{1/2}=1 \nonumber \\
& &\kappa_{3,2}+\kappa_{3,4}=\kappa_{1,1}\;\frac{\partial}{\partial \overline{V}(2,2)}(-\overline{g}_{SS})^{1/2}=-1 \nonumber
\end{eqnarray}
which fix the coefficients: $\kappa_{3,4}=\kappa_{3,3}=\kappa_{3,1}=1/2;\; \kappa_{3,2}=-3/2$.

For the case of two composite bodies  these $1/c^6$ potentials in the spatial metric component produce interbody potentials with internalized gravitational energy contributions to mass:
\begin{displaymath}
\sum_{\alpha}\kappa_{3,\alpha}\;V(3,\alpha,\vec{r}\:)\rightarrow 
\frac{G^3}{2c^6}\left\{\left(\frac{1}{R_{AB}}\left[\frac{3}{R_A}+\frac{1}{R_B}\right]-\frac{3}{R_AR_B}\right)\sum_{aa'}\frac{m_am_{a'}}{r_{aa'}}\sum_bm_b+A\leftrightarrow B
+ \frac{1}{R_A^2}\&\frac{1}{R_B^2}\;\mbox{terms}\;\right\}
\end{displaymath}
Reexpressing in proper coordinates the internal gravitational energy contributions to $1/R$ order spatial metric potentials produces some $1/R^2$ order additions to the spatial metric:
\[ 
\frac{2G\:\left(M_A-\sum_am_a\right)}{c^2\:R_A}\rightarrow -\frac{G^2}{c^4\:R_A}\frac{m_am_{a'}}{r_{aa'}}\rightarrow -\frac{G^2}{c^4\:R_A}\frac{m_am_{a'}}{\rho_{aa'}}
\left[1+\frac{G\:M_B}{c^2\:R_{AB}}\right]\;+\;A\leftrightarrow B
\]
Together, the gravitational energy contributions to composite body masses enforce interior effacement, $m_A\rightarrow M_A=\sum_a\left(m_a-\sum_{a'}Gm_{a'}/2c^2r_{aa'}+....\right)$ to the $1/c^6$ level within the spatial metric expansion.  

Altogether, the spatial metric expansion to $m^5/r^5$ order is found to be:
\begin{eqnarray}
& &-g_{SS}(\vec{r}\:)=1 + 2\sum_i\frac{G\:m_i}{c^2\:r_i}+\frac{G^2}{c^4}\sum_{ij}\left\{\frac{3}{2}\frac{m_im_j}{r_ir_j}-\frac{m_im_j}{r_ir_{ij}}\right\} \nonumber \\
& &+ \frac{G^3}{c^6}\sum_{ijk}\left\{\frac{1}{2}\frac{m_im_jm_k}{r_ir_jr_k}-\frac{3}{2}\frac{m_im_jm_k}{r_ir_jr_{jk}}
+\frac{1}{2}\frac{m_im_jm_k}{r_ir_{ij}r_{ik}}+\frac{1}{2}\frac{m_im_jm_k}{r_ir_{ij}r_{jk}}\right\} \nonumber \\
& &+\frac{G^4}{c^8}\sum_{ijkl}\left\{\frac{1}{16}\frac{m_im_jm_km_l}{r_ir_jr_kr_l}-\frac{3}{4}\frac{m_im_jm_km_l}{r_ir_jr_kr_{kl}} 
\; +\frac{3}{8}\frac{m_im_jm_km_l}{r_ir_jr_{ik}r_{jl}}+\frac{3}{4}\frac{m_im_jm_km_l}{r_ir_jr_{jk}r_{jl}}
\; +\frac{3}{4}\frac{m_im_jm_km_l}{r_ir_jr_{jk}r_{kl}} \right. \nonumber \\
& &\;\left.-\frac{1}{4}\frac{m_im_jm_km_l}{r_ir_{ij}r_{ik}r_{il}} 
\;-\frac{1}{2}\frac{m_im_jm_km_l}{r_ir_{ij}r_{ik}r_{kl}}-\frac{1}{4}\frac{m_im_jm_km_l}{r_ir_{ij}r_{jk}r_{jl}}-\frac{1}{4}\frac{\:m_im_jm_km_l}{r_ir_{ij}r_{jk}r_{kl}}\right\} \nonumber \\
& &+ \frac{G^5}{c^{10}}\sum_{ijklp}\left\{0\:\frac{m_im_jm_km_lm_p}{r_ir_jr_kr_lr_p}
\;-\frac{1}{8}\frac{m_im_jm_km_lm_p}{\:r_ir_jr_kr_lr_{lp}}
\;+\frac{3}{8}\frac{m_im_jm_km_lm_p}{r_ir_jr_kr_{jl}r_{kp}}+\frac{3}{8}\frac{m_im_jm_km_lm_p}{r_ir_jr_kr_{kl}r_{kp}} \right.\nonumber \\
& &\;+\frac{3}{8}\frac{m_im_jm_km_lm_p}{r_ir_jr_kr_{kl}r_{lp}}
\;-\frac{3}{8}\frac{m_im_jm_km_lm_p}{r_ir_jr_{jk}r_{il}r_{ip}} 
\;-\frac{3}{8}\frac{m_im_jm_km_lm_p}{r_ir_jr_{ik}r_{il}r_{ip}}
\;-\frac{3}{8}\frac{m_im_jm_km_lm_p}{r_ir_jr_{jk}r_{kl}r_{kp}} \nonumber \\
& &\;+\frac{1}{8}\frac{m_im_jm_km_lm_p}{r_ir_{ij}r_{ik}r_{il}r_{ip}}
+\frac{3}{8}\frac{m_im_jm_km_lm_p}{r_ir_{ij}r_{ik}r_{il}r_{lp}}
+\frac{1}{8}\frac{m_im_jm_km_lm_p}{r_ir_{ij}r_{ik}r_{jl}r_{kp}} 
+\frac{1}{4}\frac{m_im_jm_km_lm_p}{r_ir_{ij}r_{ik}r_{kl}r_{lp}}
+\frac{1}{4}\frac{m_im_jm_km_lm_p}{r_ir_{ij}r_{jk}r_{jl}r_{lp}} \nonumber \\
& &\;\left.+\frac{1}{4}\frac{m_im_jm_km_lm_p}{r_ir_{ij}r_{jk}r_{jl}r_{jp}}
+\frac{1}{8}\frac{m_im_jm_km_lm_p}{r_ir_{ij}r_{jk}r_{jl}r_{jp}}
+\frac{1}{8}\frac{m_im_jm_km_lm_p}{r_ir_{ij}r_{jk}r_{kl}r_{kp}}
+\frac{1}{8}\frac{m_im_jm_km_lm_p}{r_ir_{ij}r_{jk}r_{kl}r_{lp}}\right\}
\end{eqnarray}
The $\kappa_{4,\alpha}$ and $\kappa_{5,\alpha}$ coefficients were determined almost solely by use of inner effacement as described more fully in Appendix B. The $\kappa_{4,\alpha}$ results were confirmed by checking that they fulfilled all twelve of the algebraic equations for the nine $\kappa_{4,\alpha}$ shown in Appendix A and obtained from exterior effacement enforcement.    Having the spatial metric expansion to sufficiently high order, it is worthwhile confirming that interior effacement is enforced when the $1/c^8$ order potentials are internalized to the $1/c^4$ order:  the resulting potentials are
\begin{eqnarray}
& &\sum_{\alpha}\kappa_{4,\alpha}\;V(4,\alpha,\vec{r}\:)\rightarrow \frac{G^4}{c^8} 
\left\{\frac{3}{4}\left(\frac{1}{R_AR_B}-\frac{1}{R_{AB}}\left[\frac{1}{R_A}+\frac{1}{R_B}\right]\right)\sum_{aa'}\frac{m_am_{a'}}{r_{aa'}} \sum_{bb'}\frac{m_bm_{b'}}{r_{bb'}}
\right.  \\ 
& &+\left.\left[\frac{1}{2}\left(\frac{3}{R_AR_B}-\frac{1}{R_{AB}}\left[\frac{5}{R_A}+\frac{1}{R_B}\right]\right)\sum_{aa'a''}\frac{m_am_{a'}m_{aa''}}{r_{aa'}r_{aa''}} \sum_bm_b+A\leftrightarrow B\right]+ \frac{1}{R_A^2}\&\frac{1}{R_B^2}\;terms\; \right\}  \nonumber
\end{eqnarray}
The rescaling to proper internal coordinates of spatial metric's $1/R$ order composite body mass contributions from two orders of internal gravitational mass-energy produce additional $1/R^2$ order pieces to the spatial metric:
\begin{eqnarray} 
& &\frac{2G\left(M_A-\sum_am_a\right)}{c^2\:R_A}\rightarrow -\frac{G^2}{c^4\:R_A}\frac{m_am_{a'}}{r_{aa'}}+\frac{G^3}{c^6\:R_A}\frac{m_am_{a'}m_{aa''}}{r_{aa'}r_{aa''}}  \rightarrow \nonumber \\
& &\hspace{.71in}-\frac{G^2}{c^4\:R_A}\frac{m_am_{a'}}{\rho_{aa'}}\left[1+\frac{G\:M_B}{c^2\:R_{AB}}\right]
+\frac{G^3}{c^6\:R_A}\frac{m_am_{a'}m_{aa''}}{\rho_{aa'}\rho_{aa''}}\left[1+2\frac{G\:M_B}{c^2\:R_{AB}}\right] \;+\;A\leftrightarrow B
\nonumber \end{eqnarray}
Together, the Lagrangian's $V(4,\alpha)$ potentials and these rescaling contributions produce the required interior effacement of the masses appearing in the spatial metric's order $M^2$ potentials.

Confirmed now to the $1/c^8$ order of the spatial metric expansion, a conjecture is adopted: the motion-independent potentials in the isotropic portion of the spatial metric potentials fulfill to \bf all orders \rm the gravitational potential energy part of interior effacement for the spatial metric expansion. Using this conjecture, in the Appendix B iterative algebraic conditions from both exterior and interior effacement are used to obtain an analytic recursive equation for its potential coefficients $\kappa_{n,1}$ which constructs the complete spatial Schwarzschild metric potential:
\[
 2(n+1)\;\kappa_{n+1,1}=(4-n)\;\kappa_{n,1}
\]
Starting with assumed $\kappa_{1,1}=2$ it yields the finite polynomial $(1+Gm/2c^2r)^4$ with coefficient sequence 2, 3/2, 1/2, 1/16, 0,.... 0, .... for the  $\kappa_{n,1}$, $n=1,\;2,\;3,\;4,\;5,....$. This conjecture also leads to strong interrelationships between the motion independent expansion coefficients of the spatial metric, $\kappa_{n,\alpha}$ and those coefficients of the Lagrangian, $\lambda_{n,\epsilon}$.  Some examples of these interrelationships are outlined below.

1. Many of the $\kappa_{n,\alpha}$ are determinable from a known lower order coefficient by enforcing interior effacement. In the simplest case, the coefficients are inferred one successive order at a time. For any metric potential $V(n,\alpha)$ with a unique (no other equivalent) terminal site $z$,\footnote{Internalization must occur on a potential's branch terminal site.  Otherwise, rescaling to proper coordinates of a lower order, partially internalized metric potential could contribute to the relationship being here developed.} an internalized interbody interval $r_{zz'}$ can be added between site $z$ to another mass $m_{z'}$ at site $z'$. In order to enforce interior effacement of the spatial metric, the coefficient for this new metric potential $V(n+1,\alpha')$ with additional interval $r_{zz'}$ is fixed so as to produce the Newtonian potential energy contribution to composite body $M_z$ of the original potential --- $m_z\rightarrow M_Z=m_z+....-\lambda_{0,1}Gm_zm_{z'}/c^2r_{zz'}+...$. Therefore:
\begin{equation}
\kappa_{n+1,\alpha'}=-\lambda_{0,1}\;\kappa_{n,\alpha}=-\frac{1}{2}\;\kappa_{n,\alpha}\hspace{.4in}
\end{equation} 
If the initial metric potential terminal site is one of $n_i$ equivalent sites, or the final metric potential terminal site is one of $n_f$ equivalent sites, the previously derived relationship is modified to become:
\begin{displaymath}
n_f\:\kappa_{n+1,\alpha'}=-\frac{1}{2}\;n_i\;\kappa_{n,\alpha}
\end{displaymath}
This rule is illustrated by a series of spatial metric potential contributions of increasing order whose coefficients are inferred order by order:
\begin{displaymath}
\frac{1}{2}\sum_{ijk}\frac{G^3\:m_im_jm_k}{c^6\underbrace{r_ir_jr_k}_{n_i=3}}\overbrace{\longrightarrow}^{-3/2} -\frac{3}{4}\sum_{ijkl}\frac{G^4\:m_im_jm_km_l}
{c^8\:\underbrace{r_ir_j}_{n_i=2}r_kr_{kl}} 
\overbrace{\longrightarrow}^{-1/2} \frac{3}{8}\sum_{ijklp}\frac{G^5\:m_im_jm_km_lm_p}{c^{10}r_ir_jr_k\underbrace{r_{kl}r_{jp}}_{n_f=2}}
\overbrace{\longrightarrow}^{-1/6} -\frac{1}{16}\sum_{ijklpq}\frac{G^6\:m_im_jm_km_lm_pm_q}{c^{12}r_ir_jr_k\underbrace{r_{kl}r_{jp}r_{iq}}_{n_f=3}} 
\end{displaymath}  Or two orders can be added at branch terminal $z$ of the original metric potential, with two internalized intervals, $r_{zz'}$ and $r_{zz''}$ giving potential type $\alpha''$, and alternatively $r_{zz'}$ and $r_{z'z''}$ giving potential type $\alpha'''$.  The sum of these augmented potentials must produce the second order gravitational potential mass-energy for composite body $M_Z$.  This yields the iterative relationship:
\begin{displaymath}
\kappa_{n+2,\alpha''}+\kappa_{n+2,\alpha'''}=-\lambda_{1,1}\;\kappa_{n,\alpha}=\frac{1}{2}\;\kappa_{n,\alpha}
\end{displaymath}
and so on.

2. Interior effacement of the lowest order spatial metric potential $\kappa_{1,1}V(1,1)$ for a composite body source means that this body's mass requires contributions from the sum of internalized $n$th order motion-independent Lagrangian potentials 
\begin{displaymath}
m_i\longrightarrow M_i = m_i+....-\frac{1}{c^2}\sum_{\epsilon}\lambda_{n,\epsilon}\;U(n,\epsilon,int)+....
\end{displaymath}
Every Lagrangian potential $U(n,\epsilon)$ has $n+2$ masses and $n+1$ intermass intervals.  So uni-stem metric potentials $V(n+2,\alpha)$, internalized on all intervals except their stems, must produce one or the other of the completely internalized $U(n,\epsilon ,int)$. $\alpha_{\epsilon}$ are all such uni-stem metric potentials which produce a particular $U(n,\epsilon)$:
\begin{displaymath}
V(n+2,\alpha_{\epsilon})\longrightarrow \frac{G}{c^4R}\:U(n,\epsilon)
\end{displaymath}
If we know the mappings of the uni-stem metric potentials onto the Lagrangian potentials, interior effacement would require the several equations: 
\begin{displaymath}
\sum_{uni-step\;\alpha_{\epsilon}}\kappa_{n+2,\alpha_{\epsilon}}=-\kappa_{1,1}\;\lambda_{n,\epsilon}\hspace{.5in}\mbox{for each}\;\epsilon
\end{displaymath}
In absence of that knowledge of the specific mappings of the $\alpha$ onto the $\epsilon$, the overall sum rule relationship still holds for any $n$:
\begin{displaymath}
\sum_{uni-step\;\alpha}\kappa_{n+2,\alpha}=-\kappa_{1,1}\;\sum_{\epsilon}\lambda_{n,\epsilon}
\end{displaymath}

3.  Starting with any motion-independent Lagrangian potential  $\lambda_{n-1,\epsilon}\;U(n-1,\epsilon )$  which has a mass $m_z$ at a unique branch terminal location, and then comparing to another Lagrangian potential contribution $\lambda_{n,\epsilon'}\;U(n,\epsilon ')$ which differs only by having an additional interbody link between mass element $m_z$ and an additional mass element $m_{z'}$, both these mass elements are internalized to be within a composite body $Z$:
\begin{eqnarray}
U(n-1,\epsilon)&=& \sum_{ij...yz}\frac{G^n\:m_im_j....m_ym_z}{c^{2(n-1)}....r_{yz}} 
\equiv \sum_zm_z\:V(n,\alpha(\epsilon),\vec{r}_z)\nonumber \\
U(n,\epsilon')&= &\sum_{ij...yzz'}\frac{G^{n+1}\:m_im_j....m_ym_zm_{z'}}{c^{2n}....r_{yz}r_{zz'}}\simeq
\sum_{ij...y}\frac{G^n\:m_im_j....m_y}{c^{2n}....r_{yZ}}\;\sum_{zz'}\frac{G\:m_zm_{z'}}{r_{zz'}} \nonumber \\
&\simeq &V(n,\alpha,\vec{r}_Z)\;\sum_{zz'}\frac{G\:m_zm_{z'}}{r_{zz'}} \nonumber
\end{eqnarray}
$\alpha(\epsilon)$ is the type of uni-stem metric potential present when looking back into $U(n-1,\epsilon)/m_{z}$ from branch terminal location $\vec{r}_z$.
Interior effacement with regard to the composite body $Z$ at this branch terminal location then requires the energy contribution $-\lambda_{n-1,\epsilon}\;U(n-1,\epsilon )$ but with $\sum_zm_z$ replaced by $-\sum_{zz'}Gm_zm_{z'}/2c^2r_{zz'}$. This leads to the iterative equation among coefficients
\begin{equation}
\lambda_{n,\epsilon'(\epsilon )}=-\frac{1}{2}\lambda_{n-1,\epsilon}-\frac{1}{4}\kappa_{n,\alpha(\epsilon )}
\end{equation} 
The $\kappa_{n,\alpha(\epsilon )}$ term enters this equation because the potential $V(n,\alpha(\epsilon),\vec{r}_Z)$ appears in the spatial metric expansion which rescales into proper local coordinates at $\vec{r}_Z$ the Newtonian potential energy contribution of composite body $M_Z$:
\begin{displaymath}
-\frac{1}{2}\sum_{zz'}\frac{G\:m_zm_{z'}}{r_{zz'}}=-\frac{1}{2}\sum_{zz'}\frac{G\:m_zm_{z'}}{\rho_{zz'}}\;\left[-g_{SS}(\vec{r}_Z)\right]^{1/2}=
-\frac{1}{2}\sum_{zz'}\frac{G\:m_zm_{z'}}{\rho_{zz'}}\;
\left[1+....+\frac{1}{2}\kappa_{n,\alpha(\epsilon )}\;V(n,\alpha(\epsilon ),\vec{r}_Z)+....\right]
\end{displaymath}
because a uni-stem potential $V(n,\alpha(\epsilon) )$ can not be the product of two or more lower order potentials.

If Equation 18 is applied to the potentials $U(n,last)$, because of the equivalence of both ends of the potentials $U(n,last)$, the recursive equation for the associate coefficients is altered to:
\[
\lambda_{n,last}+\frac{1}{2}\lambda_{n-1,last}=-\frac{1}{8}\kappa_{n,last}=-\left(-\frac{1}{2}\right)^{n+1}
\]
This leads to the sequence of values $\lambda_{n,last}=-1/2,\;3/8,\;-1/4,\;5/32,\;.... $ for $n=1,\;2,\;3,\;4,\;....$

\subsection{The Temporal Metric Potential Expansion}

The scaling partition of the Lagrangian potential
\[
U(0,1)F(1,1)=\lambda_{0,1,1,1} \sum_{ij}\frac{G\:m_im_j}{r_{ij}}\;\frac{v_{ij}^2}{c^2} \longrightarrow 2\lambda_{0,1,1,1}\;\overline{V}(1,1)\;m_i v_i^2
\]
when equated to the required scaling of the Newtonian kinetic energy $m_iv_i^2\:/2$
\[
2\lambda_{0,1,1,1}=\frac{1}{2}\frac{\partial}{\partial\:\overline{V}(1,1)}(\overline{g}_{00})^{-1/2}(-\overline{g}_{SS})^1=3/2
\]
fixes $\lambda_{0,1,1,1}=3/4$.  The iterative constraints related to time dilation and Lorentz contraction then fix the rest of the $L_{m^2v^2/r}$ coefficients:
\begin{equation}
L_{m^2v^2/r}=\frac{1}{4}\frac{G\:m_im_j}{c^2\;r_{ij}}\left(3\:v_{ij}^2-\vec{v}_i\cdot\vec{v}_j-\vec{v}_i \cdot\hat{r}_{ij}\hat{r}_{ij}\cdot \vec{v}_j\right)
\end{equation}
and the test particle limit of this potential when equated to the geodesic Lagrangian  requires the temporal metric potential $1/c^4$ contribution:
\[
\delta g_{00}=-3\frac{G\:m_i}{c^4\;r_i}\:v_i^2
\]
This motion-dependent potential along with the potential $\xi_{2,2}\;V(2,2,r_{ij}...),\;\xi_{2,2}=2$ in $g_{00}$ show that the temporal metric potential's Newtonian level mass parameter for a composite body consists not solely of that body's mass-energy, but in addition contains a scalar virial term shown in brackets $[....]$:
\begin{equation}
M^*=\sum_i m_i\left(1 +\frac{1}{2c^2}\left(v_i^2-\sum_j\frac{G\:m_j}{r_{ij}}\right)+....\right)+\frac{1}{c^2}\sum_im_i\left[v_i^2-\frac{1}{2}\sum_j\frac{G\:m_j}{r_{ij}}+....\right]
\end{equation}

Continuing the specification of the Lagrangian and $g_{00}$ expansions to next order, the scaling partitions of the Lagrangian's three motion-independent potentials at the next order are:
\begin{eqnarray}
& &U(3,1)\rightarrow 4\overline{V}(1,1)\;U(2,1)+6\overline{V}(2,1)\;U(1,1)+4\overline{V}(3,1)\;U(0,1) \nonumber \\
& &U(3,2)\rightarrow \overline{V}(1,1)\;U(2,1)+2 \overline{V}(1,1)\;U(2,2) \nonumber \\
& &\hspace{.61in}+\left[2\overline{V}(2,1)+\overline{V}(2,2)\right]U(1,1)+\left[\overline{V}(3,1)+2\overline{V}(3,2)+\overline{V}(3,3)\right]U(0,1) \nonumber \\
& &U(3,3)\rightarrow 2\overline{V}(1,1)\;U(2,2)+\left[2\overline{V}(2,2)+\overline{V}(2,1)\right]U(1,1)+2\left[\overline{V}(3,2)+\overline{V}(3,4)\;\right]U(0,1)
\nonumber 
\end{eqnarray}
They support eight exterior effacement algebraic conditions from Equation 8:
\begin{eqnarray}
& &4\lambda_{3,1}+\lambda_{3,2}=\lambda_{2,1}\;\frac{\partial}{\partial \overline{V}(1,1)}(\overline{g}_{00})^{1/2}(-\overline{g}_{SS})^{-3/2}=-1 \nonumber \\
& &2\lambda_{3,2}+2\lambda_{3,3}=\lambda_{2,2}\;\frac{\partial}{\partial \overline{V}(1,1)}(\overline{g}_{00})^{1/2}(-\overline{g}_{SS})^{-3/2}=-3/2 \nonumber \\
& &6\lambda_{3,1}+3\lambda_{3,2}+\lambda_{3,3}=\lambda_{1,1}\;\frac{\partial}{\partial \overline{V}(2,1)}(\overline{g}_{00})^{1/2}(-\overline{g}_{SS})^{-1}=-5/2 \nonumber \\
& &\lambda_{3,2}+2\lambda_{3,3}=\lambda_{1,1}\;\frac{\partial}{\partial \overline{V}(2,2)}(\overline{g}_{00})^{1/2}(-\overline{g}_{SS})^{-1}=-1 \nonumber \\
& &4\lambda_{3,1}+\lambda_{3,2}=\lambda_{0,1}\;\frac{\partial}{\partial \overline{V}(3,1)}(\overline{g}_{00})^{1/2}(-\overline{g}_{SS})^{-1/2}=-1 \nonumber \\
& &2\lambda_{3,2}+2\lambda_{3,3}=\lambda_{0,1}\;\frac{\partial}{\partial \overline{V}(3,2)}(\overline{g}_{00})^{1/2}(-\overline{g}_{SS})^{-1/2}=-3/2 \nonumber \\
& &\lambda_{3,2}=\lambda_{0,1}\;\frac{\partial}{\partial \overline{V}(3,3)}(\overline{g}_{00})^{1/2}(-\overline{g}_{SS})^{-1/2}=-1/2  \nonumber \\
& &2\lambda_{3,3}=\lambda_{0,1}\;\frac{\partial}{\partial \overline{V}(3,4)}(\overline{g}_{00})^{1/2}(-\overline{g}_{SS})^{-1/2}=-1/2 \nonumber 
\end{eqnarray}
with solutions $\lambda_{3,1}=-1/8,\;\lambda_{3,2}=-1/2,\;\lambda_{3,3}=-1/4$.   Carrying out specification of the Lagrangian to one more order using interior and exterior effacement produces altogether: 
\begin{eqnarray}
& &L=\frac{1}{2}\sum_{ij} \frac{G\: m_i m_j}{r_{ij}} - \frac{1}{2}\sum_{ijk} \frac{G^2\: m_i m_jm_k}{c^2\:r_{ij}r_{ik}}
+\frac{G^3}{c^4}\sum_{ijkl}\left\{\frac{1}{4} \frac{ m_i m_jm_km_l}{r_{ij}r_{ik}r_{il}}+\frac{3}{8} \frac{ m_i m_jm_km_l}{r_{ij}r_{jk}r_{kl}}\right\} \nonumber \\
& &-\frac{G^4}{c^6}\sum_{ijklp}\left\{\frac{1}{8} \frac{ m_i m_jm_km_lm_p}{r_{ij}r_{ik}r_{il}r_{ip}}
+\frac{1}{2} \frac{ m_i m_jm_km_lm_p}{r_{ij}r_{ik}r_{il}r_{lp}}
+\frac{1}{4} \frac{ m_i m_jm_km_lm_p}{r_{ij}r_{jk}r_{kl}r_{lp}}\right\} \nonumber \\
& &+\frac{G^5}{c^8}\sum_{ijklpq}\left\{\frac{1}{16} \frac{m_i m_jm_km_lm_pm_q}{r_{ij}r_{ik}r_{il}r_{ip}r_{iq}}
+\frac{5}{16} \frac{ m_i m_jm_km_lm_pm_q}{r_{ij}r_{ik}r_{il}r_{ip}r_{pq}} 
+\frac{5}{32} \frac{ m_i m_jm_km_lm_pm_q}{r_{ij}r_{ik}r_{il}r_{jp}r_{jq}} \right.  \nonumber \\
& &\hspace{.4in}+\left.\frac{5}{16} \frac{ m_i m_jm_km_lm_pm_q}{r_{ij}r_{jk}r_{kl}r_{lp}r_{lq}} 
+\frac{5}{16} \frac{ m_i m_jm_km_lm_pm_q}{r_{ij}r_{ik}r_{il}r_{kp}r_{lq}}+\frac{5}{32} \frac{ m_i m_jm_km_lm_pm_q}{r_{ij}r_{jk}r_{kl}r_{lp}r_{pq}}\right\} \nonumber
\end{eqnarray}

The scaling partitions for the temporal metric's four potentials $V(3,\alpha)$ are:
\begin{eqnarray}
& &V(3,1)\longrightarrow  3\overline{V}(1,1)\;V(2,1) + 3\overline{V}(2,1)\;V(1,1) \nonumber \\
& &V(3,2)\longrightarrow \overline{V}(1,1)\;\left\{V(2,1)+V(2,2)\right\}+\left[\overline{V}(2,1)+\overline{V}(2,2)\right]\;V(1,1) \nonumber \\
& &V(3,3)\longrightarrow 2\overline{V}(1,1)\;V(2,2) + \overline{V}(2,1)\;V(1,1) \nonumber \\
& &V(3,4)\longrightarrow \overline{V}(1,1)\;V(2,2) + \overline{V}(2,2)\;V(1,1) \nonumber
\end{eqnarray}
and they support four exterior effacement algebraic conditions from Equation 7:
\begin{eqnarray}
& &3\xi_{3,1}+\xi_{3,2} = \xi_{2,1}\;\frac{\partial}{\partial \bar{V}(1,1)}(\bar{g}_{00})^1(-\bar{g}_{SS})^{-1}   =-8 \nonumber \\
& &\xi_{3,2}+2\xi_{3,3}+\xi_{3,4}=\xi_{2,2}\;\frac{\partial}{\partial \bar{V}(1,1)}(\bar{g}_{00})^1(-\bar{g}_{SS})^{-1} =-8 \nonumber \\
& &3\xi_{3,1}+\xi_{3,2}+\xi_{3,3}=\xi_{1,1}\;\frac{\partial}{\partial \bar{V}(2,1)}(\bar{g}_{00})^1(-\bar{g}_{SS})^{-1/2}=-19/2 \nonumber \\
& &\xi_{3,2}+\xi_{3,4}=\xi_{1,1}\;\frac{\partial}{\partial \bar{V}(2,2)}(\bar{g}_{00})^1(-\bar{g}_{SS})^{-1/2}=-5 \nonumber
\end{eqnarray}
Equating the test particle limit of the Lagrangian expansion to the geodesic Lagrangian fixes $\xi_{3,1}=-3/2$. Then the above algebraic constraints from enforcing exterior effacement fix the rest of the coefficients to be $\xi_{3,2}=-7/2,\;\xi_{3,3}=\xi_{3,4}=-3/2$.  
\begin{eqnarray}
& &g_{00}(\vec{r}\:)=1 - 2\sum_i\frac{G\:m_i}{c^2\:r_i}+\frac{G^2}{c^4}\sum_{ij}\left\{2\frac{m_im_j}{r_ir_j}+2\frac{\:m_im_j}{\:r_ir_{ij}}\right\} \nonumber \\ 
& &-\frac{G^3}{c^6}\sum_{ijk}\left\{\frac{3}{2}\frac{m_im_jm_k}{r_ir_jr_k} 
+\frac{7}{2}\frac{m_im_jm_k}{r_ir_jr_{jk}}
+\frac{3}{2}\frac{m_im_jm_k}{r_ir_{ij}r_{ik}}+\frac{3}{2}\frac{m_im_jm_k}{r_ir_{ij}r_{jk}}\right\} \nonumber \\ 
& &+\frac{G^4}{c^8}\sum_{ijkl}\left\{\frac{m_im_jm_km_l}{r_ir_jr_kr_l}+\frac{7}{2}\frac{m_im_jm_km_l}{r_ir_jr_kr_{kl}}+....+\xi_{4,9}\;\frac{m_im_jm_km_l}{r_ir_{ij}r_{jk}r_{kl}}\right\}
\end{eqnarray}
The derivation of the full Schwarzschild temporal metric potential 
\[
\sqrt{g_{00}(r)}=\frac{1-Gm/2c^2r}{1+Gm/2c^2r}=1-\frac{Gm}{c^2r}+\frac{1}{2}\left(\frac{Gm}{c^2r}\right)^2-\frac{1}{4}\left(\frac{Gm}{c^2r}\right)^3+....
\]
using a combination of interior and exterior effacement enforcement is given in the Appendix B.

\section{The Lagrangian's Motion-Dependent Potentials}
The lowest order motion-dependent Lagrangian potentials 
\[ 
L_{m^2v^2/r}=\frac{1}{4}\frac{G\:m_im_j}{c^2\;r_{ij}}\left(3\:v_{ij}^2-\vec{v}_i\cdot\vec{v}_j-\vec{v}_i \cdot\hat{r}_{ij}\hat{r}_{ij}\cdot \vec{v}_j\right)
\]
have already been determined, Equation 19, because they were needed to establish an interesting property of the composite body mass parameters that appear in the temporal metric.  The $1/c^4$ order velocity-dependent Lagrangian potentials take two forms:
\begin{eqnarray}
& &L_{m^2v^4/r}=\sum_{ij}\frac{G\;m_i m_j }{c^4\:r_{ij}}
\left\{ \lambda_{0,1,2,1} \:v_{ij}^4 +\lambda_{0,1,2,2} \: v_{ij} ^2 \;\vec{v}_i \cdot \vec{v}_j 
+ \lambda_{0,1,2,3}\:v_i ^2 v_j ^2 +\lambda_{0,1,2,4} \: (\vec{v}_i \cdot \vec{v}_j )^2  \right. \nonumber\\
 & & + \left[ \lambda_{0,1,2,5} \:  v_{ij} ^2  +\lambda_{0,1,2,6} \: \vec{v}_i \cdot \vec{v}_j + \lambda_{0,1,2,7} \: \vec{v}_i \cdot \hat{r}_{ij}\hat{r}_{ij} 
\cdot \vec{v}_j  +
\lambda_{0,1,2,8} (( \hat{r}_{ij} \cdot \vec{v}_i )^2  + ( \hat{r}_{ij} \cdot \vec{v}_j )^2 ) \right]\; \vec{v}_i \cdot \hat{r}_{ij} \hat{r}_{ij} \cdot \vec{v}_j  \nonumber \\
& &\left. +\:  \lambda_{0,1,2,9} \left[ ( v_i ^2 - \vec{v}_i \cdot \vec{v}_j ) ( \hat{r}_{ij} \cdot \vec{v}_j )^2
+ ( v_j ^2 - \vec{v}_i \cdot \vec{v}_j) ( \hat{r}_{ij} \cdot \vec{v}_i )^2    \right] \right\} 
\end{eqnarray}
\begin{eqnarray}
& &L_{m^3v^2/r^2} = \sum_{ijk}\frac{G^2\:m_i m_j m_k}{c^4\:r_{ij} r_{ik}}
\left\{ \lambda_{1,1,1,1}\: v_i ^2 + \lambda_{1,1,1,2} \:\left( v_j ^2 + v_k ^2 \right) +  \lambda_{1,1,1,3}\: \vec{v}_i \cdot \left( \vec{v}_j + \vec{v}_k \right)
+ \lambda_{1,1,1,4}\: \vec{v}_j \cdot \vec{v}_k  \right. \nonumber \\
& & \left. + \lambda_{1,1,1,5}\: \vec{v}_i \cdot \left( \hat{r}_{ij} \hat{r}_{ij} \cdot \vec{v}_j + \hat{r}_{ik} \hat{r}_{ik} \cdot \vec{v}_k \right)
+\lambda_{1,1,1,6}\;\vec{v}_j\cdot\left(\hat{r}_{ij} \hat{r}_{ij}  + \hat{r}_{ik} \hat{r}_{ik} \right)\cdot\vec{v}_k  \right\}
 \end{eqnarray}

The $L_{m^2v^4/r}$ potentials produce time dilation scalings of the lower order Lagrangian potentials at both the $w^4$ and $w^2$ orders 
\begin{eqnarray}
& &L_{m^2v^4/r} \Longrightarrow \left( \lambda_{0,1,2,3} +  \lambda_{0,1,2,4} \right)\; \frac{w^4}{c^4}\;\sum_{ij}\frac{G\;m_i m_j }{r_{ij}} \nonumber \\
&  &+ \frac{w^2}{c^2}\;\sum_{ij}\frac{G\;m_i m_j }{c^2\:r_{ij}} \left(\lambda_{0,1,2,2}\: v_{ij}^2 +\lambda_{0,1,2,3} \left(v_i ^2 + v_j ^2 \right)
+  2 \lambda_{0,1,2,4} \:\vec{v}_i \cdot \vec{v}_j +  \lambda_{0,1,2,6} \:\vec{v}_i \cdot \hat{r}_{ij} \hat{r}_{ij} \cdot \vec{v}_j \right) \nonumber
\end{eqnarray}
which gives three algebraic conditions for the $\lambda_{0,1,2,\tilde{\epsilon}}$ coefficients;
\[
\lambda_{0,1,2,2}+\lambda_{0,1,2,3}=3/8,\hspace{.51in} \lambda_{0,1,2,3} +\lambda_{0,1,2,4} =-1/16,\hspace{.51in}\lambda_{0,1,2,6}=-1/8 
\]
And there is also an order $\bar{V}(1,1)$ spectator scaling partition: 
\[
L_{m^2v^4/r} \Longrightarrow 2 \lambda_{0,1,2,1}\: \bar{V}(1,1)\: \sum_i m_i v_i^4 
\]
which establishes $\lambda_{0,1,2,1} = 7/16$. The Lorentz contraction condition on $L_{m^2v^4/r}$ establishes one more novel constraint for the coefficients, $\lambda_{0,1,2,7}+2\lambda_{0,1,2,8}=3/16$, which results from the scalings:
\[
L_{m^2v^4/r}\longrightarrow \left\{\lambda_{0,1,2,6}\frac{w^2\left(\vec{w}\cdot\hat{r}_{ij}\right)^2}{c^4}+\left(\lambda_{0,1,2,7}+2\lambda_{0,1,2,8}\right)\frac{\left(\vec{w}\cdot\hat{r}_{ij}\right)^4}{c^4}\right\}
\sum_{ij}\frac{G\;m_i m_j }{r_{ij}} 
\]
 
The $L_{m^3v^2/r^2}$ potentials are lean in time dilation conditions but rich in spectator scaling conditions. The time dilation scaling is
\[
L_{m^3v^2/r^2}\;\Longrightarrow \; \left(\lambda_{1,1,1,1} + 2 \lambda_{1,1,1,2} +2 \lambda_{1,1,1,3}+ \lambda_{1,1,1,4} \right)\frac{w^2}{c^2} \sum_{ijk}\frac{G^2\;m_i m_j m_k}{c^2\:r_{ij} r_{ik}} \nonumber
\]
Equating this scaling from $L_{m^3v^2/r^2}$ to the required scaling at the  $w^2$ order for the $L_{1,1}$ potential yields 
\[
\lambda_{1,1,1,1} + 2 \lambda_{1,1,1,2} +2 \lambda_{1,1,1,3}+ \lambda_{1,1,1,4} = \frac{1}{4}
\]
There is one Lorentz contraction scaling:
\[
L_{m^3v^2/r^2}\;\Longrightarrow \left(\lambda_{1,1,1,5}+\lambda_{1,1,1,6}\right)\frac{\vec{w}\cdot(\hat{r}_{ij}\hat{r}_{ij}+\hat{r}_{ik}\hat{r}_{ik})\cdot\vec{w}}{c^2}
\sum_{ijk}\frac{G^2\;m_i m_j m_k}{c^2\:r_{ij} r_{ik}}
\]
And $L_{m^3v^2/r^2}$ produces five spectator scalings of the two lower order Lagrangian potentials:
\begin{eqnarray}
L\;\Longrightarrow&\;& \left(\lambda_{1,1,1,1}\overline{V}(2,1) + 2 \lambda_{1,1,1,2} \overline{V}(2,2) \right) \sum_i m_i v_i ^2   \nonumber     \\
&+&\overline{V}(1,1)\;\sum_{ij}\frac{G\;m_i m_j }{c^2\:r_{ij} }\left( [\lambda_{1,1,1,1} +  \lambda_{1,1,1,2} ][ v_i ^2 + v_j ^2 ] + 2\lambda_{1,1,1,3} \vec{v}_i \cdot  \vec{v}_j
 + 2 \lambda_{1,1,1,5} \vec{v}_i \cdot  \hat{r}_{ij} \hat{r}_{ij} \cdot \vec{v}_j \right)  \nonumber
\end{eqnarray}
The required spectator scaling factors for these partitions are:
\[
\frac{\partial}{\partial \overline{V}(2,1)}(-\overline{g}_{SS})^1 (\overline{g}_{00})^{-1/2}=4 \hspace{.3in}
 \frac{\partial}{\partial \overline{V}(2,2)}(-\overline{g}_{SS})^1 (\overline{g}_{00})^{-1/2} =-2 \hspace{.3in}
\frac{\partial}{\partial \overline{V}(1,1)}(-\overline{g}_{SS})^{1/2} (\overline{g}_{00})^{-1/2}=2 
\]
Comparing the explicit scaling contributions from $L_{m^3v^2/r^2}$  to the required scalings for lower order potentials yields solutions for all of its coefficients:
\[
\lambda_{1,1,1,1} =2,\hspace{.2in} \lambda_{1,1,1,2} =-1/2,\hspace{.2in} \lambda_{1,1,1,3} = -7/4,\hspace{.2in} 
\lambda_{1,1,1,4} = 11/4,\hspace{.2in} \lambda_{1,1,1,5} =-1/4\hspace{.2in}\lambda_{1,1,1,6}=1/2 
\]

\subsection{Newtonian $1/R$ Energy Between Bodies' Internal Kinetic Energies.}

Interior effacement requires that the $1/R_{AB}$ potential energy between two composite bodies be enforced internal energy types by internal energy types, since they result from the internalizations of  different fundamental Lagrangian potential types .  The kinetic energies of two composite bodies contribute to the overall potential energy:
\[
\delta E = -\frac{G}{c^4\:R_{AB}}\sum_a\frac{1}{2}m_av_a^2\sum_b\frac{1}{2}m_bv_b^2
\]
But rescalings  of composite body internal kinetic energies due to metric potentials of the other body yield $-4G\:\sum_am_av_a^2\:\sum_bm_bv_b^2\:/c^4R_{AB}$:
\[
\frac{1}{2}\left( \sum_am_av_a^2+\sum_bm_bv_b^2\right) \rightarrow 
\frac{1}{2}\sum_am_au_a^2\; \left[1-\frac{2G(M_B^*+M_B)}{c^2\:R_{AB}}\right]+\frac{1}{2}\sum_bm_bu_b^2\; \left[1-\frac{2G(M_A^*+M_A)}{c^2\:R_{AB}}\right]
\]
$L_{m^2v^4/r}$ produces a variety of energy contributions of this type in amounts
\begin{eqnarray}
& &E_{m^2v^4/r}=\sum_i\vec{v}_i\cdot\frac{\partial L_{m^2v^4/r} }{\partial\vec{v}_i}-L_{m^2v^4/r}  =3L_{m^2v^4/r}\rightarrow \sum_{ab}\frac{G\:m_am_b}{c^4\:R_{AB}}\left\{
\left(12\lambda_{0,1,2,1}+6\lambda_{0,1,2,3}\right)\;v_a^2v_b^2 \right. \nonumber \\
& &+\left(24\lambda_{0,1,2,1}-12\lambda_{0,1,2,2}+6\lambda_{0,1,2,4}\right)\left(\vec{v}_a\cdot\vec{v}_b\right)^2 -\left(12\lambda_{0,1,2,5}-6\lambda_{0,1,2,6}\right)\;\vec{v}_a\cdot\vec{v}_b\:\vec{v}_a\cdot\hat{R}\hat{R}\cdot\vec{v}_b \nonumber \\
& &\left. +6\lambda_{0,1,2,7}\;\left(\vec{v}_a\cdot\hat{R}\hat{R}\cdot\vec{v}_b\right)^2
+6\lambda_{0,1,2,9}\;\left(v_a^2\left(\hat{R}\cdot\vec{v}_b\right)^2+v_b^2\left(\hat{R}\cdot\vec{v}_a\right)^2 \right) \right\}
\end{eqnarray}
If there were no more contributions to energies of this type, interior effacement would require  
\begin{eqnarray}
15/4& &= 12\lambda_{0,1,2,1}+6\lambda_{0,1,2,3} \nonumber \\
0& &=24\lambda_{0,1,2,1}-12\lambda_{0,1,2,2}+6\lambda_{0,1,2,4}=12\lambda_{0,1,2,5}-6\lambda_{0,1,2,6}=6\lambda_{0,1,2,7}=6\lambda_{0,1,2,9} \nonumber
\end{eqnarray}
But these algebraic constraints are inconsistent with the previously fixed values among the $\lambda_{0,1,2,\tilde{\epsilon}}$ coefficients. This points to the need for acceleration-dependent Lagrangian potentials  at this $1/c^4$ order:
\begin{eqnarray}
& &L_{m^2v^2a}= \sum_{ij}\frac{G\;m_i m_j }{c^4\:r_{ij}}\left\{ \lambda_{0,1,2,10}\left[\left( v_j^2-\vec{v}_i\cdot\vec{v}_j\right)\: \vec{a}_i
-\left(v_i^2-\vec{v}_i\cdot\vec{v}_j\right)\:\vec{a}_j\right] \right. \nonumber \\
& & +\lambda_{0,1,2,11} \: \left(\vec{a}_i\cdot\vec{v}_j\vec{v}_j  -\vec{a}_j\cdot \vec{v}_i\vec{v}_i\right)
 +\lambda_{0,1,1,12}\:\vec{a}_{ij}\cdot\left(\vec{v}_i\vec{v}_j+\vec{v}_j\vec{v}_i \right)+\lambda_{0,1,2,13}\:(\vec{a}_i+\vec{a}_j)\cdot(\vec{v}_i\vec{v}_j-\vec{v}_j\vec{v}_i) \nonumber \\  
& & \left.  + \lambda_{0,1,2,14}\:[(\vec{v}_j\cdot\hat{r}_{ij})^2\:\vec{a}_i
- (\vec{v}_i\cdot\hat{r}_{ij})^2  \:\vec{a}_j]+\lambda_{0,1,2,15}\: \vec{v}_i\cdot\hat{r}_{ij}\vec{v}_j\cdot\hat{r}_{ij}\:\vec{a}_{ij}    
\right\}\cdot\vec{r}_{ij} 
\end{eqnarray}
The $1/c^2$ order Lagrangian contains no acceleration-dependent potentials, so this general expansion for the $1/c^4$ order acceleration-dependent Lagrangian $L_{m^2v^2a}$ has been explicitly constructed so as to show no spectator or time dilation scalings of a lower order acceleration-dependent Lagrangian potential. With the energy contributions from Lagrangians dependent on both velocity and acceleration being:
\[
E= \sum_i \left(\vec{v}_i \cdot \frac{\partial L}{\partial \vec{v}_i}- L+\vec{a}_i \cdot \frac{\partial L}{\partial \vec{a}_i}- \vec{v}_i
\cdot \frac{d}{dt}\frac{\partial L}{\partial \vec{a}_i}  \right)
\]
The yield of energies quartic in velocities from  $L_{m^2v^2a}$ are 
\[
E= -\sum_{ik} \vec{v}_i\cdot\frac{\partial^2L_{m^2v^2a}}{\partial\vec{a}_i\:\partial\vec{r}_k}   \cdot\vec{v}_k
\]
Adding contributions from the acceleration-dependent Lagrangian, the five equations for $1/c^4$ coefficients then become:
\begin{eqnarray}
& &12\lambda_{0,1,2,1}+6\lambda_{0,1,2,3}-4\lambda_{0,1,2,10}+4\lambda_{0,1,2,12}+4\lambda_{0,1,2,13}=15/4  \nonumber \\
& &24\lambda_{0,1,2,1}-12\lambda_{0,1,2,2}+6\lambda_{0,1,2,4}-4\lambda_{0,1,2,10}-4\lambda_{0,1,2,11}+4\lambda_{0,1,2,12}-4\lambda_{0,1,2,13}= 0 \nonumber \\
& &-12\lambda_{0,1,2,5}+6\lambda_{0,1,2,6}+4\lambda_{0,1,2,10}+4\lambda_{0,1,2,11}-4\lambda_{0,1,2,12}+4\lambda_{0,1,2,13}-8\lambda_{0,1,2,14}+8\lambda_{0,1,2,15}= 0 \nonumber \\
& &6\lambda_{0,1,2,7}+12\lambda_{0,1,2,14}-12\lambda_{0,1,2,15}=0 \nonumber \\
& &6\lambda_{0,1,2,9}+2\lambda_{0,1,2,10}-2\lambda_{0,1,2,12}-2\lambda_{0,1,2,13}-2\lambda_{0,1,2,14}+2\lambda_{0,1,2,15}=0 \nonumber
\end{eqnarray}

\subsection{Newtonian $1/R$ Energy Between Bodies' Internal Kinetic and Gravitational Energies.}

Interior effacement of the Newtonian energy between two composite bodies also requires the contribution
\[  \delta E = \frac{G}{c^4\;R_{AB}}\left(\sum_a\frac{1}{2}m_au_a^2\;\sum_{bb'}\frac{G\:m_bm_{b'}}{2\:r_{bb'}}
                                         +\sum_b\frac{1}{2}m_bu_b^2\;\sum_{aa'}\frac{G\:m_am_{a'}}{2\:r_{aa'}}\right)  \]
The $L_{m^3v^2/r}$ potential with partial internalization contributes energies of this type to the system with coefficient $2\lambda_{1,1,1,2}$
and the rescaling to proper coordinates also produces energies of this same form with overall coefficient $5/4$.  For the previously fixed $\lambda_{1,1,1,2}=-1/2$ the net coefficient for this energy is the $1/4$ value required for interior effacement. This requires three constraints on coefficients from the acceleration-dependent
$L_{m^2v^2a}$ potential such that no further energies of this type are produced from that Lagrangian.
\begin{eqnarray}
& &2\lambda_{0,1,2,10}+\lambda_{0,1,2,12}+\lambda_{0,1,2,13}=0 \nonumber \\
& &\lambda_{0,1,2,10}-2\lambda_{0,1,2,11}-\lambda_{0,1,2,12}+\lambda_{0,1,2,13}=0 \nonumber \\
& &2\lambda_{0,1,2,14}+\lambda_{0,1,2,15}=0  \nonumber
\end{eqnarray}
The eight algebraic equations from this and previous subsection have the solution
\begin{eqnarray} 
& &\lambda_{0,1,2,7}=\lambda_{0,1,2,9}=\lambda_{0,1,2,10}=\lambda_{0,1,2,14}=\lambda_{0,1,2,15}=0 \nonumber \\ 
& &\lambda_{0,1,2,11}=\lambda_{0,1,2,13}=-\lambda_{0,1,2,12}=11/32  \nonumber
\end{eqnarray}
This solution also enforces interior effacement to the $1/c^4$ order for the $1/R_{AB}^2$ potential energy between composite bodies $A$ and $B$.

\subsection{ Newtonian $1/R$ Energy Between Bodies' Internal Gravitational Energies.}

Interior effacement of the Newtonian energy between two composite bodies also requires the contribution
\[
\delta E = -\frac{G}{c^4\;R_{AB}}\sum_{bb'}\frac{G\:m_bm_{b'}}{2\:r_{bb'}}\sum_{aa'}\frac{G\:m_am_{a'}}{2\:r_{aa'}}
\]
The Lagrangian potential $U(2,2)$ by partial internalization contributes energy of this form to the system
\[
E_{2,2}=-2\lambda_{2,2}\;\frac{G}{c^4\;R_{AB}}\sum_{bb'}\frac{G\:m_bm_{b'}}{r_{bb'}}\sum_{aa'}\frac{G\:m_am_{a'}}{r_{aa'}}
\]
Rescalings to proper coordinates also produce energies of this same form with numerical coefficient $1/2$.  Noting the previously fixed $\lambda_{2,2}=3/8$, the net coefficient for this energy is the $-1/4$ value required for interior effacement.

All coefficients for the $1/c^4$ order N-body Lagrangian have now been fixed:
\begin{eqnarray}
&L &=-\sum_i m_i \left(c^2-\frac{1}{2}v_i^2-\frac{1}{8c^2}v_i^4-\frac{1}{16c^4}v_i^6\:\right) \nonumber \\
&+&\frac{1}{2}\sum_{ij}\frac{G\;m_im_j}{r_{ij}}-\frac{1}{2}\sum_{ijk}\frac{G^2\:m_im_jm_k}{c^2\:r_{ij}r_{ik}}+\frac{1}{4}\sum_{ijkl}\frac{G^3\:m_im_jm_km_l}{c^4\:r_{ij}r_{ik}r_{il}}
+\frac{3}{8}\sum_{ijkl}\frac{G^3\:m_im_jm_km_l}{c^4\:r_{ij}r_{ik}r_{kl}}  \nonumber \\
&+& \frac{1}{4}\sum_{ij}\frac{G\:m_i m_j }{c^2\:r_{ij}} \left\{3\:  v_{ij} ^2  -\:\vec{v}_i \cdot \vec{v}_j
-\vec{v}_i \cdot  \hat{r}_{ij} \hat{r}_{ij} \cdot \vec{v}_j \right\} \nonumber \\
&+&\sum_{ijk}\frac{G^2\:m_i m_j m_k}{c^4\:r_{ij} r_{ik}}
 \left\{ 2\: v_i ^2 -\frac{1}{2}\: \left( v_j ^2 + v_k ^2 \right) -\frac{7}{4}\: \vec{v}_i \cdot \left( \vec{v}_j + \vec{v}_k \right)+ \frac{11}{4} \vec{v}_j \cdot \vec{v}_k \right. \nonumber \\ &- &\left. \frac{1}{4}\: \vec{v}_i \cdot \left( \hat{r}_{ij} \hat{r}_{ij} \cdot \vec{v}_j + \hat{r}_{ik} \hat{r}_{ik} \cdot \vec{v}_k \right) 
+\frac{1}{2} \vec{v}_j\cdot\left(\hat{r}_{ij}\hat{r}_{ij}+\hat{r}_{ik}\hat{r}_{ik}\right)\cdot \vec{v}_k                        \right\} \nonumber \\
&+&\sum_{ij}\frac{G\;m_i m_j }{c^4\:r_{ij}}
\left\{ \frac{7}{16} \:v_{ij}^4 +\frac{5}{8} \: v_{ij} ^2 \;\vec{v}_i \cdot \vec{v}_j 
-\frac{1}{4}\:v_i ^2 v_j ^2 +\frac{3}{16} \: (\vec{v}_i \cdot \vec{v}_j )^2  \right. \nonumber \\
&+& \left. \left[ \frac{9}{32} \:  v_{ij} ^2  -\frac{1}{8} \: \vec{v}_i \cdot \vec{v}_j  +\frac{3}{32}
\left(( \hat{r}_{ij} \cdot \vec{v}_i )^2  + ( \hat{r}_{ij} \cdot \vec{v}_j )^2 \right) \right]\; \vec{v}_i \cdot \hat{r}_{ij} \hat{r}_{ij} \cdot \vec{v}_j
         \right\} \nonumber \\
&+& \sum_{ij}\frac{G\;m_i m_j }{c^4\:r_{ij}}
\left\{\frac{11}{32}\left(  \vec{a}_i\cdot\vec{v}_j\left[\vec{v}_j-2\vec{v}_i\right]  -\vec{a}_j\cdot \vec{v}_i\left[\vec{v}_i-2\vec{v}_j\right] \right) \right\}\cdot\vec{r}_{ij} 
\end{eqnarray}
except for one undetermined coefficient of a total time derivative 
\[
\frac{dQ}{dt}=\frac{d}{dt}\sum_{ij}\frac{Gm_im_j}{c^4\:r_{ij}^3}\left[\left(\vec{v}_j\cdot\vec{r}_{ij}\right)^2\vec{v}_i\cdot\vec{r}_{ij}
-\left(\vec{v}_i\cdot\vec{r}_{ij}\right)^2\vec{v}_j\cdot\vec{r}_{ij}\right]
\]
contained within the original general Lagrangian $L_{m^2v^4/r}+L_{m^2v^2a}$.

\appendix
\section{Iterative Algebraic Equations for Spatial Metric's  Coefficients}

Twelve algebraic conditions for nine spatial metric expansion coefficients, $\kappa_{4,\alpha},\;\alpha=1\;to\;9$, result from the totality of their spectator scaling partitions, and then the application of the iterative scaling requirements for exterior effacement given by Equation 6:
\begin{eqnarray}
& &4\kappa_{4,1}+\kappa_{4,2}=\kappa_{3,1}\;\frac{\partial}{\partial \overline{V}(1,1)}(-\overline{g}_{SS})^{-1/2}=-1/2 \nonumber \\
& &2\kappa_{4,2}+2\kappa_{4,3}+2\kappa_{4,4}+\kappa_{4,5}=\kappa_{3,2}\;\frac{\partial}{\partial \overline{V}(1,1)}(-\overline{g}_{SS})^{-1/2}=3/2 \nonumber \\
& &\kappa_{4,4}+3\kappa_{4,6}+\kappa_{4,7}=\kappa_{3,3}\;\frac{\partial}{\partial \overline{V}(1,1)}(-\overline{g}_{SS})^{-1/2}=-1/2 \nonumber \\
& &\kappa_{4,5}+\kappa_{4,7}+2\kappa_{4,8}+\kappa_{4,9}=\kappa_{3,4}\;\frac{\partial}{\partial \overline{V}(1,1)}(-\overline{g}_{SS})^{-1/2}=-1/2 \nonumber \\
& &6\kappa_{4,1}+2\kappa_{4,2}+\kappa_{4,3}+\kappa_{4,4}=\kappa_{2,1}\;\frac{\partial}{\partial \overline{V}(2,1)}(-\overline{g}_{SS})^0=0 \nonumber \\
& &\kappa_{4,2}+\kappa_{4,5}=\kappa_{2,1}\;\frac{\partial}{\partial \overline{V}(2,2)}(-\overline{g}_{SS})^0=0 \nonumber \\
& &\kappa_{4,2}+2\kappa_{4,4}+\kappa_{4,5}+3\kappa_{4,6}+\kappa_{4,7}+\kappa_{4,8}=\kappa_{2,2}\;\frac{\partial}{\partial \overline{V}(2,1)}(-\overline{g}_{SS})^0=0 \nonumber \\
& &2\kappa_{4,3}+\kappa_{4,7}+\kappa_{4,9}=\kappa_{2,2}\;\frac{\partial}{\partial \overline{V}(2,2)}(-\overline{g}_{SS})^0=0 \nonumber \\
& &4\kappa_{4,1}+\kappa_{4,2}+\kappa_{4,4}+\kappa_{4,6}=\kappa_{1,1}\;\frac{\partial}{\partial \overline{V}(3,1)}(-\overline{g}_{SS})^{1/2}=0  \nonumber \\
& &2\kappa_{4,2}+\kappa_{4,3}+\kappa_{4,5}+\kappa_{4,7}=\kappa_{1,1}\;\frac{\partial}{\partial \overline{V}(3,2)}(-\overline{g}_{SS})^{1/2}=-1/2  \nonumber \\
& &\kappa_{4,4}+\kappa_{4,8}=\kappa_{1,1}\;\frac{\partial}{\partial \overline{V}(3,3)}(-\overline{g}_{SS})^{1/2}=1/2  \nonumber \\
& &\kappa_{4,5}+\kappa_{4,9}=\kappa_{1,1}\;\frac{\partial}{\partial \overline{V}(3,4)}(-\overline{g}_{SS})^{1/2}=1/2  \nonumber
\end{eqnarray}

All of these twelve equations required by exterior effacement are fulfilled by the coefficients shown in Equation 15, although those coefficient values were first obtained  by use of the lowest order interior effacement requirement for composite body masses as represented by the substitutions 
\begin{displaymath}
m_i\longrightarrow M_I=\sum_i\left(m_i+....-\sum_{i'}Gm_{i'}/2c^2r_{ii'}+....\right)
\end{displaymath}
which indicates some of the redundancy from the two types of effacement.  The corresponding twelve algebraic conditions for the nine temporal metric expansion coefficients are constructed by making the substitutions $\kappa_{4,\alpha}\longrightarrow \xi_{4,\alpha}$ on left hand sides of the twelve equations, and modifying the right hand side expressions for the required scaling factors --- $(-\overline{g}_{SS})^x\longrightarrow (\overline{g}_{00})^1(-\overline{g}_{SS})^{x-1}$.

\section{Obtaining The Complete Spatial and Temporal Schwarzschild Metric Potentials}

Using a combination of exterior and interior effacement, the spatial metric's first two potential coefficients at all orders can be fixed by iterative algebraic equations.\footnote{In the single source case, the potentials $V(n,1)$ are the only potentials to survive.  And collectively, the $\kappa_{n,1}$ and $\xi_{n,1}$ coefficients build the complete spatial and temporal Schwarzschild metric potentials, respectively.}  Focusing on the potentials:
\[
V(n+1,1)=\sum_{i....n+1}\frac{G^{n+1}\:m_i....m_nm_{n+1}}{c^{2(n+1)}r_ir_j....r_nr_{n+1}}\hspace{.5in}
V(n+1,2)=\sum_{i....n+1} \frac{G^{n+1}\:m_i....m_nm_{n+1}}{c^{2(n+1)}r_ir_j....r_n\:r_{n,n+1}}
\]
the lowest order $\overline{V}(1,1)$ spectator scaling potential may be factored out by partitioning, leaving local potentials $V(n,1)$ of one lower order:
\begin{eqnarray}
& &\kappa_{n+1,1}\;V(n+1,1)\longrightarrow (n+1)\;\kappa_{n+1,1}\;\overline{V}(1,1)\;V(n,1) \nonumber \\
& &\kappa_{n+1,2}\;V(n+1,2)\longrightarrow\kappa_{n+1,2}\; \overline{V}(1,1)\;V(n,1) \nonumber
\end{eqnarray}
Exterior effacement then requires for all $n$
\[
(n+1)\:\kappa_{n+1,1}+\kappa_{n+1,2}= \kappa_{n,1} \;\frac{\partial}{\partial \overline{V}(1,1)}(-\overline{g}_{SS})^{1-n/2}=(2-n)\kappa_{n,1}
\]
Interior effacement requires that the potential $\kappa_{n+1,2}\;V(n+1,2)$, internalized on its $r_{n,n+1}$ interval, yields the Newtonian gravitational energy contributions for all the composite body masses on the $n$ equivalent branch terminal locations of $V(n,1)$:
\[
\kappa_{n+1,2}=-\frac{n}{2}\;\kappa_{n,1}
\]
Together, these exterior and interior effacement requirements result in the recursive equation
\[
 2(n+1)\;\kappa_{n+1,1}=(4-n)\;\kappa_{n,1}
\]
which generates the finite polynomial Schwarzschild spatial metric potential $(1+Gm/2c^2r)^4$ coefficient sequence 2, 3/2, 1/2, 1/16, 0,.... 0, .... for the  $\kappa_{n,1}$, $n=1,\;2,\;3,\;4,\;5,....$.  And then also the coefficients $\kappa_{n,2}$ are fixed as $-1,\;-3/2,\;-3/4,\;-1/8,\;0,....0,....$ for $n=2,\;3,\;4,\;5,\;6,....$

In order to obtain the complete temporal Schwarzschild metric potential consider the two metric potentials
\begin{displaymath}
V(n+1,st[2])\sim \frac{1}{\underbrace{r_{jk}r_{jl}....r_{jz}}_{n-1} \underbrace{r_ir_{ij}}_2}\hspace{.4in}
V(n+1,st[3])\sim \frac{1}{\underbrace{r_{kl}r_{km}....r_{jz}}_{n-2} \underbrace{r_ir_{ij}r_{jk}}_3}
\end{displaymath}
Interior effacement enforcement for the spatial metric potentials $V(n,last)$ then yields for any $n$ the series of algebraic equations among coefficients:
\begin{eqnarray}
& &\kappa_{n+1,st[1]}+\kappa_{n+1,st[2]}=-\kappa_{1,last}\;\lambda_{n-1,1} \nonumber \\
& &\kappa_{n+1,st[2]}+\kappa_{n+1,st[3]}=-\kappa_{2,last}\;\lambda_{n-2,1} \nonumber \\
& &\kappa_{n+1,st[3]}+\kappa_{n+1,st[4]}=-\kappa_{3,last}\;\lambda_{n-3,1} \nonumber \\
& &\kappa_{n+1,st[4]}+\kappa_{n+1,st[5]}=-\kappa_{4,last}\;\lambda_{n-4,1} \nonumber \\
& &.... \nonumber \\
& &\kappa_{n+1,st[n-1]}+\kappa_{n+1,st[n]}=-\kappa_{n-1,last}\;\lambda_{1,1} \nonumber
\end{eqnarray}
Recalling that $\kappa_{n,last}=-4(-1/2)^n$ for all $n$, this set of equations has the general solution
\[
\lambda_{n,1}=(-1/2)^n\; \mbox{for}\; n>0
\]
along with the $\kappa_{n,st[n']}$  being independent of stem length $n'$ for any given $n$.  
The derived iterative equation 
\[(n+1)\lambda_{n,1}=-\frac{n+2}{2}\;\lambda_{n-1,1}+\frac{1}{4}\;\kappa_{n,st[1]} \]
with $\kappa_{n,st[1]}=-4(-1/2)^n$ also confirms the above solution for $\lambda_{n,1}$.
Taking the test particle limit of the Lagrangian and equating it to the geodesic Lagrangian then yields  the Schwarzschild temporal metric potential: 
\[
\sqrt{g_{00}(r)}=\frac{1-Gm/2c^2r}{1+Gm/2c^2r}=1-\frac{Gm}{c^2r}+\frac{1}{2}\left(\frac{Gm}{c^2r}\right)^2-\frac{1}{4}\left(\frac{Gm}{c^2r}\right)^3+....
\]

\subsection{ Can Effacements Be Enforced With $\kappa_{1,1} \neq 2?$}

This appendix assumes starting point for spatial metric expansion to be $-g_{SS}(\vec{r}\:)=1+2\gamma \: U(\vec{r}\:)/c^2$, and then examines whether a Lagrangian expansion to order $1/c^4$ and its accompanying metric field expansions can be found which enforce exterior and interior effacements for general $\gamma$ value?
The construction done in the body of the paper is (successfully) redone here in abbreviated form for general $\gamma$.

Requiring exterior effacement for the Newtonian potential energy at lowest order
\[ 2\lambda_{1,1}= \lambda_{0,1}\;\frac{\partial}{\partial \overline{V}(1,1)}(\overline{g}_{00})^{1/2}(-\overline{g}_{SS})^{-1/2}=-\frac{\gamma+1}{2} \]
Newton's $G$ then remains at this lowest order locally unaffected by distribution of distant matter.\cite{N84}. Proceeding to next order gives some modified partitions
\begin{eqnarray}
& & V(2,1)=\sum_{ij}\frac{G^2\:m_im_j}{c^4\;r_ir_j}\longrightarrow 2\overline{V}(1,1)\;\sum_i \frac{G\:m_i}{c^2\:r_i}= 2 \overline{V}(1,1)\;V(1,1) \nonumber \\
& & V(2,2)= \sum_{ij}\frac{G^2\:m_im_j}{c^4\;r_ir_{ij}}\longrightarrow\overline{V}(1,1)\;V(1,1) \nonumber
\end{eqnarray}
which support two algebraic equations for metric  potential coefficients:
\begin{eqnarray}
& & 2\xi_{2,1}+\xi_{2,2}= \xi_{1,1}\frac{\partial}{\partial\:\overline{V}(1,1)}(\overline{g}_{00})(-\overline{g}_{SS})^{-1/2}=-(2+\gamma)\xi_{1,1}=2(2+\gamma) \nonumber \\ & & 2\kappa_{2,1}+\kappa_{2,2}=\kappa_{1,1}\frac{\partial}{\partial\:\overline{V}(1,1)}(-\overline{g}_{SS})^{1/2}=2\;\gamma^2 \nonumber 
\end{eqnarray}
Taking the single test particle limit of the Lagrangian and equating it to a geodesic equation then fixes the temporal metric potential coefficients $\xi_{2,1}=(3+\gamma)/2$ and $\xi_{2,2}=1+\gamma$. The fixing of the spatial metric coefficients $\kappa_{2,\alpha}$ requires a combined use of exterior effacement and interior effacement. Two $1/c^4$ order motion-independent potentials in the N-body Lagrangian have multiple scaling partitions:
\begin{eqnarray}
U(2,1)=\sum_{ijkl}\frac{G^3\:m_im_jm_km_l}{c^4\:r_{ij}r_{ik}r_{il}}&\longrightarrow& 3\overline{V}(1,1)\;U(1,1)+3\overline{V}(2,1)\;U(0,1) \nonumber \\
U(2,2)=\sum_{ijkl}\frac{G^3\:m_im_jm_km_l}{c^4\:r_{ij}r_{jk}r_{kl}}&\longrightarrow&  2\overline{V}(1,1)\;U(1,1)+\left[\overline{V}(2,1)+2\overline{V}(2,2)\right]\;U(0,1)
\nonumber \end{eqnarray}
which support three algebraic equations  for $\lambda(2,1)$ and $\lambda(2,2)$:
\begin{eqnarray}
3\lambda_{2,1}+2\lambda_{2,2}&=&\lambda_{1,1}\;\frac{\partial}{\partial \overline{V}(1,1)}\left((\overline{g}_{00})^{1/2}(-\overline{g}_{SS})^{-1}\right)=
(1+\gamma)(1+2\gamma)/4 \nonumber \\
3\lambda_{2,1}+\lambda_{2,2}&=&\lambda_{0,1}\;\frac{\partial}{\partial \overline{V}(2,1)}\left((\overline{g}_{00})^{1/2}(-\overline{g}_{SS})^{-1/2}\right)
=\left([3\gamma+1][2\gamma+1]-2\kappa_{2,1}\right)/8 \nonumber \\
2\lambda_{2,2}&=&\lambda_{0,1}\frac{\partial}{\partial \overline{V}(2,2)}\left((\overline{g}_{00})^{1/2}(-\overline{g}_{SS})^{-1/2}\right)=\left(1+\gamma-\kappa_{2,2}\right)/4
\nonumber \end{eqnarray}
For  two composite bodies A and B at rest and separated by $R$, the Lagrangian potential $\lambda_{2,2}\;U(2,2)$ includes a $1/R$ energy contribution:
\[
-\lambda_{2,2}\:U(2,2)\longrightarrow-2\lambda_{2,2}\;\frac{G}{R}\sum_{aa'}\frac{G\;m_am_{a'}}{c^2\:r_{aa'}}\sum_{bb'}\frac{G\;m_bm_{b'}}{c^2\:r_{bb'}}
\]
but interior effacement requires the total coefficient of such form to be $-1/4$. Energy contributions of this same form from the rescaling of the bodies' Newtonian gravitational energies into their proper spatial coordinates give:
\begin{displaymath}
-\frac{1}{2}\left(\sum_{aa'}\frac{G\;m_am_{a'}}{c^2\:r_{aa'}}+\sum_{bb'}\frac{G\;m_bm_{b'}}{c^2\:r_{bb'}}\right)\rightarrow -\frac{1}{2}\sum_{aa'}\frac{G\;m_am_{a'}}{c^2\:\rho_{aa'}}\left[1+\gamma\:\frac{G\:M_B}{c^2\:R}\right]-\frac{1}{2}\sum_{bb'}\frac{G\;m_bm_{b'}}{c^2\:\rho_{bb'}}\left[1+\gamma\:\frac{G\:M_A}{c^2\:R}\right]
\end{displaymath}
The coefficient of the Newtonian gravitational energy to $M_B$ in the spatial metric component is $\kappa_{2,2}\:/2\gamma$.  So altogether interior effacement requires $\lambda_{2,2}=(1-2\kappa_{2,2})/8$.  With the previous equation for $\lambda_{2,2}$ obtained from an exterior effacement condition, there is:
\[
\lambda_{2,2}=\left(1+\gamma-\kappa_{2,2}\right)/8=(1-2\kappa_{2,2})/8
\]     
with solution $\kappa_{2,2}=-\gamma$ and $\kappa_{2,1}=\gamma(\gamma+2)/2$. Then $\lambda_{2,2}=(1+2\gamma)/8$ and $\lambda_{2,1}=\gamma(1+2\gamma)/12$.  The composite body mass parameter $M$ appearing at lowest order in the spatial metric component expansion remains having the "normal" multiple  of its Newtonian gravitational energy contribution $-G\sum_{aa'}m_am_{a'}/2c^2\:r_{aa'}$. The $\gamma$-dependent metric field potential expansions are now:
\begin{eqnarray}
& &g_{00}=1-2V(1,1)+ \frac{\gamma+3}{2}V(2,1) + (\gamma+1)V(2,2)+ .... \nonumber \\
& &-g_{SS}=1+2\gamma V(1,1) +\frac{\gamma(\gamma+2)}{2}V(2,1)-\gamma V(2,2) + .... \nonumber
\end{eqnarray}
Three parts of the Lagrangian directly interrelated by exterior spectator scaling are now considered to see if their coefficients can be fixed to fulfill exterior effacement?
\begin{eqnarray}
& &L_{mv^2}+L_{m^2v^2/r}+L_{m^3v^2/r^2}=\frac{1}{2}\sum_im_iv_i^2+\lambda_{0,1,1,1}\;\sum_{ij}\frac{G\:m_i m_j }{c^2\:r_{ij}} \left\{v_i^2+v_j^2  \right\} \nonumber \\
& &\hspace{.51in}+\sum_{ijk}\frac{G^2\:m_i m_j m_k }{c^4\:r_{ij}r_{ik}}\left\{\lambda_{1,1,1,1}v_i^2+\lambda_{1,1,1,2}\left(v_j^2+v_k^2\right)\right\} \nonumber
\end{eqnarray}
For $\gamma=1$ the coefficients needed for exterior effacement are $\lambda_{0,1,1,1}=3/4,\;\lambda_{1,1,1,1}=2,\;\lambda_{1,1,1,2}=-1/2$. The pertinent scaling partitions for the case of general $\gamma$ are: 
\begin{eqnarray}
& &L_{m^2v^2/r}= \lambda_{0,1,1,1}\;\sum_{ij}\frac{G\:m_i m_j }{c^2\:r_{ij}} \left\{v_i^2+v_j^2\right\} + ....\rightarrow 2\lambda_{0,1,1,1}\;\overline{V}(1,1)\sum_im_iv_i^2 \nonumber \\
& & L_{m^3v^2/r^2}=\sum_{ijk}\frac{G^2\:m_i m_j m_k }{c^4\:r_{ij}r_{ik}}\left\{\lambda_{1,1,1,1}v_i^2+\lambda_{1,1,1,2}\left(v_j^2+v_k^2\right)\right\}+....\rightarrow \nonumber \\
& &\hspace{.3in}(\lambda_{1,1,1,1}+\lambda_{1,1,1,2}\overline{V}(1,1)\;\sum_{ij}\frac{G\:m_i m_j }{c^2\:r_{ij}} \left\{v_i^2+v_j^2 \right\}
+[\lambda_{1,1,1,1}\;\overline{V}(2,1)+2\lambda_{1,1,1,2}\;\overline{V}(2,2)]\;\sum_im_iv_i^2 \nonumber  
\end{eqnarray}
And the pertinent extractions from the required scaling factors are:
\[ \frac{\partial}{\partial \overline{V}(2,1)}(-\overline{g}_{SS})^1 (\overline{g}_{00})^{-1/2}=\frac{4\gamma^2+9\gamma+3}{4} \hspace{1in}
 \frac{\partial}{\partial \overline{V}(2,2)}(-\overline{g}_{SS})^1 (\overline{g}_{00})^{-1/2} =-\frac{3\gamma+1}{2} \]
\[ \frac{\partial}{\partial \overline{V}(1,1)}(-\overline{g}_{SS})^{1/2} (\overline{g}_{00})^{-1/2}=(1+\gamma) 
\hspace{1in} \frac{\partial}{\partial \overline{V}(1,1)}(-\overline{g}_{SS})^1 (\overline{g}_{00})^{-1/2}=(1+2\gamma) \]
For exterior effacement the $\overline{V}(1,1)$ level scaling of the Newtonian kinetic energy Lagrangian requires $\lambda_{0,1,1,1}=(1+2\gamma)/4$.  That then fulfills interior effacement to the $1/c^2$ order for the Newtonian potential energy between composite bodies $A$ and $B$:
\[ \sum_{ab}m_am_b \longrightarrow \sum_{ab}\left( m_am_b+\left[ \frac{m_av_a^2}{2c^2}-\sum_{a'}\frac{G\:m_am_{a'}}{2c^2\:r_{aa'}} \right]\:m_b
+A \leftrightarrow B \right) \]
and the temporal metric potential's Newtonian mass for a composite body is now
\[
M^*=\sum_i m_i\left(1 +\frac{1}{2c^2}\left(v_i^2-\sum_j\frac{G\:m_j}{r_{ij}}\right)+....\right)+\gamma\frac{1}{c^2}\sum_im_i\left[v_i^2-\frac{1}{2}\sum_j\frac{G\:m_j}{r_{ij}}+....\right]
\]
The $\overline{V}(2,1)$ level scaling of the Newtonian kinetic energy  requires $\lambda_{1,1,1,1}=(4\gamma^2+9\gamma+3)/8$, and the 
$\overline{V}(2,2)$ level scaling of the Newtonian kinetic energy term requires $\lambda_{1,1,1,2}=-(3\gamma+1)/8$.  And these last two coefficient values are consistent with the $\overline{V}(1,1)$ level scaling of the $1/c^2$ motion-dependent Lagrangian potential which requires $\lambda_{1,1,1,1}+\lambda_{1,1,1,2}=(2\gamma+1)(1+\gamma)/4$.   The required numerical coefficient for the internalized kinetic times gravitational energies in bodies A and B is 1/4.  The motion-independent Lagrangian potential $U(1,1,1,2)$ with its $\gamma$-dependent coefficient $\lambda_{1,1,1,2}$ contributes part of this energy, and with the rescaling contribution of this same form having total coefficient $(2+3\gamma)/4$, interior effacement is fulfilled. Only the first of the  eight equations needed to fulfill interior effacement for the kinetic times kinetic internal energies of bodies $A$ and $B$ is modified: 
\[12\lambda_{0,1,2,1}+6\lambda_{0,1,2,3}-4\lambda_{0,1,2,10}+4\lambda_{0,1,2,12}+4\lambda_{0,1,2,13}=(12\gamma+3)/4 \]
Changing the coefficient of the acceleration-dependent Lagrangian in Equation 26 from $11/32$ to $8\gamma+3)/32$ then results in interior and exterior effacement being fulfilled to the $1/c^4$ Lagrangian order for general $\gamma\neq 1$.

\end{document}